\documentclass[conference]{IEEEtran}
\IEEEoverridecommandlockouts
\usepackage{cite}
\usepackage{amsmath,amssymb,amsfonts}
\usepackage{algorithmic}
\usepackage{multirow}
\usepackage{graphicx}
\usepackage{textcomp}
\usepackage{xcolor}
\usepackage{caption}
\usepackage{url}
\usepackage{subcaption}
\newtheorem{definition}{Definition}[section]

\newcommand{\modelname}{MetaKRec }

\def\BibTeX{{\rm B\kern-.05em{\sc i\kern-.025em b}\kern-.08em
    T\kern-.1667em\lower.7ex\hbox{E}\kern-.125emX}}
\begin{document}

\title{MetaKRec: Collaborative Meta-Knowledge Enhanced Recommender System}

\author{\IEEEauthorblockN{Liangwei Yang$^{1*\thanks{$^*$The first two authors contributed equally.}}$, Shen Wang$^{1*}$, Jibing Gong$^{2}$, Shaojie Zheng$^{2}$, Shuying Du$^{2}$, Zhiwei Liu$^{3}$ and Philip S. Yu$^{1}$}
\IEEEauthorblockA{\textit{$^{1}$Department of Computer Science, University of Illinois at Chicago}, USA \\
\{lyang84, swang224, psyu\}@uic.edu\\
\textit{$^{2}$Yanshan University}, China\\
gongjibing@ysu.edu.cn, zhengshaojie@stumail.ysu.edu.cn, dsying2022@163.com \\
\textit{$^{3}$Salesforce AI Research}, USA, 
zhiweiliu@salesforce.com
}
}

\maketitle

\begin{abstract}
Knowledge graph (KG) enhanced recommendation has demonstrated improved performance in the recommendation system (RecSys) and attracted considerable research interest. Recently the literature has adopted neural graph networks (GNNs) on the collaborative knowledge graph and built an end-to-end KG-enhanced RecSys.
However, the majority of these approaches have three limitations: (1) treat the collaborative knowledge graph as a homogeneous graph and overlook the highly heterogeneous relationships among items, (2) lack of design to explicitly leverage the rich side information, and (3) overlook the rich knowledge in user preference.

To fill this gap, in this paper, we explore the rich, heterogeneous relationship among items and propose a new KG-enhanced recommendation model called Collaborative Meta-Knowledge Enhanced Recommender System (MetaKRec). In particular, we focus on modeling the rich, heterogeneous semantic relationships among items and construct several collaborative Meta-KGs to explicitly depict the relatedness of the items under the guidance of meta-knowledge. 
In addition to the knowledge obtained from KG, we leverage user knowledge that extracts from user preference to construct the Meta-KGs. The constructed Meta-KGs can capture the knowledge from both the knowledge graph and user preference.
Furthermore. we utilize a light convolution encoder to recursively integrate the item relationship in each collaborative Meta-KGs. This scheme allows us to explicitly gather the heterogeneous semantic relationships among items and encode them into the representations of items. In addition, we propose channel attention to fuse the item and user representations from different Meta-KGs. Extensive experiments are conducted on four real-world benchmark datasets, demonstrating significant gains over the state-of-the-art baselines on both regular and cold-start recommendation settings. 
\end{abstract}


\section{Introduction}
\begin{figure}
     \includegraphics[width=0.5\textwidth]{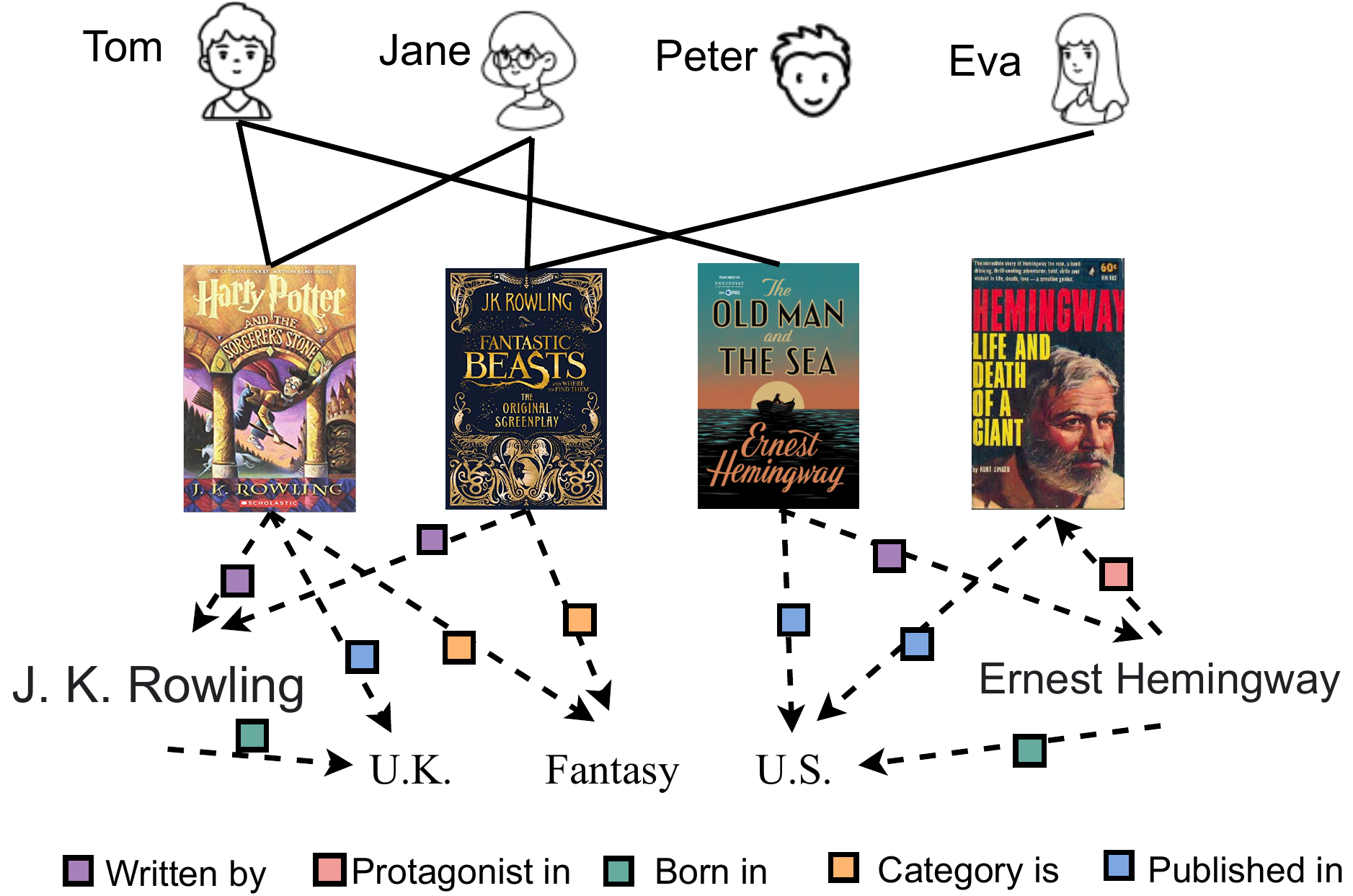}
     \caption{Illustative example on knowledge graph enhanced recommender system}
     \label{fig:illustration}
\end{figure}

Knowledge graph (KG) enhanced recommendation is becoming increasingly popular due to its significant performance gains over traditional recommendation. The benefit mainly results from better user/item representation learned with knowledge graph side information, since that side information is able to alleviate cold-start issues in the recommendation.
We illustrate the book recommendation task as a toy example in Figure~\ref{fig:illustration}: Tom bought the books ``Harry Potter" and ``The Old Man and The SEA". Without the knowledge graph information, it is not clear if ``Hemingway: Life and Death of a Giant" should be recommended to Tom. By taking the information in the knowledge graph, we know that ``Hemingway: Life and Death of a Giant" depict the life of ``Ernest Hemingway" who authors the ``The Old Man and The SEA". We can recommend ``Hemingway: Life and Death of a Giant" to Tom based because he may be interested in the life of ``Ernest Hemingway". In addition, taking the KG into consideration also brings explainability, for example, the reason ``Harry Potter" and ``Fantastic Beasts" are similar is that they are written by the same author. Therefore study Knowledge graph (KG) enhanced recommendation is a nontrivial research question.

The key to KG-enhanced recommendation is to effectively incorporate the KG-side information for learning qualitative user/item representation.
 Existing works can be classified into two categories: feature-engineered approaches and end-to-end approaches.
Feature-engineered approaches benefit from prior knowledge learned from analytic work and explicitly model the KG side information.
In earlier research~\cite{ai2018learning,cao2019unifying,ma2019jointly}, KG triplets are used to create embeddings, which are then used to enrich item representations by treating them as prior or content knowledge. To better characterization of user-item connections, several follow-up research~\cite{wang2019explainable, xian2019reinforcement, hu2018leveraging} extend the interactions with multi-hop edges from user to item. However, feature-engineered approaches usually require heavy human efforts to obtain good features.
On the other hand, the end-to-end approaches~\cite{kgat, kgin,wang2019knowledge, wang2020ckan} aim to implicitly model the KG side information by building end-to-end trained graph neural networks (GNNs) \cite{kipf2016semi,hamilton2017inductive,velivckovic2017graph} on the constructed Collaborative Knowledge Graph (CKG). The basic idea is to utilize the information aggregation strategy, which can successfully capture multi-hop structures and encode them into representations. However, the majority of these approaches have some limitations: (1) treat the collaborative knowledge graph as
a homogeneous graph and overlook the highly heterogeneous relationship among items; 
(2) lack of design to explicitly leverage the rich side information. 
(3) overlook the rich knowledge in user preference.
Thus it requires finding a solution that can overcome these limitations.

The collaborative Knowledge Graph (CKG) contains abundant heterogeneous relationships between items. Two items can be linked together through different relationships, providing fruitful information to enhance the recommender system. An illustration is shown in Figure~\ref{fig:illustration}. 
Similar items can be linked to the same entity by different kinds of relationships, e.g., we can observe two triples: (``Old Man and The Sea", ``written by", ``Ernest Hemingway") and (``Ernest Hemingway",``Protagonist in",``Hemingway: Life and Death of a Giant"). 
The two books are linked to ``Hemingway" by different relationships. Users who purchased ``Old Man and The Sea" will likely be interested in ``Hemingway: Life and Death of a Giant" because it depicts the life of the author of ``Old Man and The Sea".
Two items that share the same relationship with one entity tend to be similar. E.g., the book ``Harry Potter" and ``Fantastic Beasts" are both linked to the fantasy category. Tom, who purchased ``Harry Potter", is also likely to be interested in ``Fantastic Beasts" which belongs to the same category. 
Items can also be similar if they share similar semantic information over the knowledge graph even if they are not directly linked, e.g., ``Old Man and The Sea" does not have a link to ``Harry Potter". But they are also similar in the sense that they are both the author's most popular books. The unlinked entity similarity can be captured by the knowledge graph embedding method.
Item similarity can also be revealed by users' co-purchase behavior. E.g., Eva may be interested in ``Harry Potter" because it is co-purchased with ``Fantastic Beasts" by Jane. Thus, we can find similar items by the high co-purchase similarity or Top K co-purchased items.
We can also alleviate the cold-start problem from the relational links in the knowledge graph. E.g., we can recommend the non-interacted ``Hemingway: Life and Death of a Giant" to Tom because it shows the life of the author of ``Old Man and The Sea". Thus, it requires incorporating these heterogeneous relationships into the representation of learning of the items.

Based on the above motivation, we propose the Collaborative Meta-Knowledge Graph Enhanced Recommender System (\modelname), a novel KG-enhanced recommendation model that explores the rich heterogeneity link between items. To represent the relatedness of the items under the guidance of the meta-graph, we build a variety of collaborative Meta-KGs with a specific emphasis on modeling the rich, diverse semantic connections among items. Additionally, we develop a simple convolution encoder to recursively incorporate the item connection in every shared Meta-KG. With the help of this technique, we can explicitly collect and encode the many semantic links between things. We also propose channel attention to combine the user and item representations from several Meta-KGs. Extensive experiments on four real-world benchmark datasets show considerable improvements (to be added) over the state-of-the-art baselines for both regular and cold-start recommendation settings. We open sourced \modelname at \textcolor{blue}{\url{https://github.com/YangLiangwei/MetaKRec}}.

The key contributions are summarized as follows:
\begin{itemize}
    \item We first put forward Collaborative Meta-Knowledge Graphs to explicitly encode prior Meta-knowledge as edges between items. It is flexible to transform from both knowledge graphs and historical interactions.
    \item We propose channel attention that can fuse information from different Collaborative Meta-Knowledge Graphs. It is effective to achieve better performance by combining different Collaborative Meta-Knowledge Graphs.
    \item We propose \modelname, and validate its effectiveness on $4$ real-world datasets. It achieves the best performance under both normal and cold-start settings.
\end{itemize}

\section{Preliminaries}
This section gives the required preliminaries for \modelname, including the problem formulation and graph neural network.

\subsection{Problem Formulation}
For a knowledge graph enhanced recommender system, we have a set of users $\mathcal{U} = \{u_1,u_2,...,u_{\left | \mathcal{U} \right|}\}$, a set of items $\mathcal{I} = \{i_1,i_2,...,i_{\left | \mathcal{I} \right|}\}$, and a set of entities $\mathcal{E} = \{e_1,e_2,...,e_{\left | \mathcal{E} \right|}\}$. Historical user-item interactions are represented as a user-item bipartite graph $\mathcal{G}_{rec} = \{(u, y_{ui}, i) | u \in \mathcal{U}, i \in \mathcal{I}\}$, where $y_{ui} = 1$ if $u$ has purchased $i$, otherwise $y_{ui} = 0$. Besides $\mathcal{G}_{rec}$, we also have a knowledge graph $\mathcal{G}_{kg}$ as side information for items. The knowledge graph is typically formed by entities $\mathcal{E}$ and the relationships $\mathcal{R}$ among them. $\mathcal{G}_{kg}$ is represented by subject-property-object triple facts~\cite{kg}: $\{(h,r,t)|h,t \in \mathcal{E}, r \in \mathcal{R}\}$, where each triple represents the relation between $h$ and $t$ is $r$.

Collaborative Knowledge Graph $\mathcal{G}$~\cite{kgat} is built by unifying $\mathcal{G}_{rec}$ and $\mathcal{G}_{kg}$. Firstly, we set the item-entity alignment function $f(i)=e$ that maps item $i$ as entity $e$ in the knowledge graph. Then user-item interaction $(u, y_{ui}, i) \in \mathcal{G}_{rec}$ is transformed into two triples $(u, \text{Interact}, f(i))$ and $(f(i), \text{Interacted by}, u)$. Based on the previous alignment, the collaborative knowledge graph is represented by $\mathcal{G}=\{(h,r,t)|h,t\in \mathcal{U}\cup \mathcal{I}\cup \mathcal{E}, r \in \mathcal{R} \cup \{\text{Interact}, \text{Interacted by}\}\}$. Then the task is formulated to predict the adoption probability $\hat{y}_{ui}$ given $\mathcal{G}$ that contains both historical interactions and the knowledge graph.

\begin{figure*}
     \centering
     \includegraphics[width=0.8\textwidth]{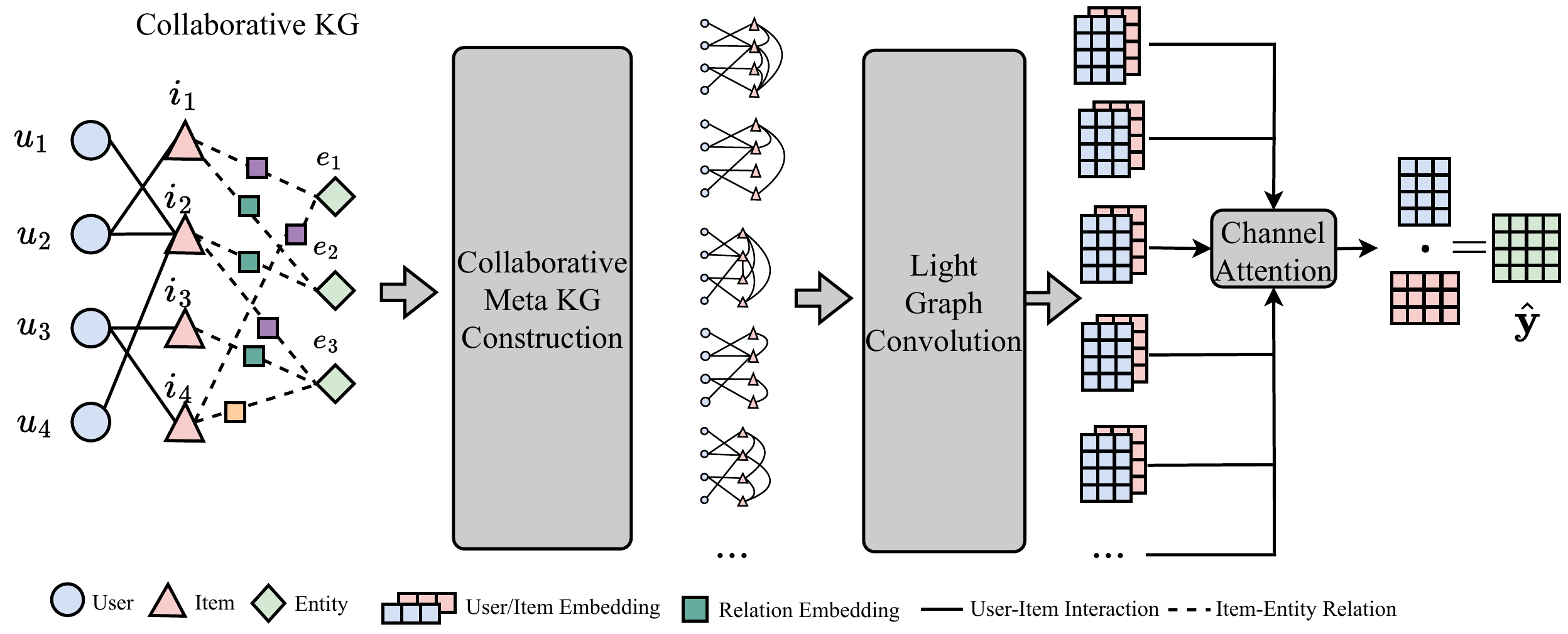}
     \caption{Framework of \modelname. The collaborative KG is fed into the Collaborative Meta KG Construction module to generate collaborative meta-kgs based on meta-knowledge. User/item embedding for each collaborative meta-kgs is then obtained by Light Graph Convolution. The Channel Attention module is finally used to fuse information over all embedding tables.}
     \label{fig:framework}
\end{figure*}

\subsection{Graph Neural Network}
Graph neural network (GNN) is a deep learning method applied to graph-structured data. It has been tested to be effective in a wide range of graph-related tasks such as protein function prediction~\cite{DeepGraphGo}, group identification~\cite{group} as well as recommender system~\cite{yang2021consisrec,liu2021federated,yang2022large}. GNN is based on the homophily assumption~\cite{newman2018networks} on graphs. It indicates nodes connected in a graph are similar to each other. GNN models homophily by directly aggregating embedding from neighbors to the center neighbor, which is formulated as:
\begin{align}
    \mathbf{e}_{u}^{(l+1)} = \mathbf{e}_{u}^{(l)} \oplus \text{AGG}^{(l+1)} (\{ \mathbf{e}_{i}^{(l)} \mid i \in \mathcal{N}_u \}),
\end{align}
where $\mathbf{e}_{u}^{(l)}$ is node $u$'s embedding in $l$-th layer, $\mathcal{N}_u$ is $u$'s neighbors, $\mbox{AGG}^{(l)}(\cdot)$ aggregates neighbors' embedding into a single vector in layer $l$, and $\oplus$ is the function to combine neighborhood representation and the center node's embedding. Different selection of $\mbox{AGG}^{(l)}(\cdot)$ and $\oplus$ leads to different kinds of GNN layers, such as GCN~\cite{gcn}, GAT~\cite{gat}, and GIN~\cite{gin}.

\section{Method}

This section presents the proposed \modelname, which includes $4$ parts. 1) Collaborative Meta-KG construction based on $\mathcal{G}_{kg}$. 2) Graph convolution on the Collaborative Meta-KGs. 3) Channel attention module aggregating embedding learned from different meta-kg channels, and 4) Prediction module to predict the probability based on learned embedding. The main framework is also shown in Fig.~\ref{fig:framework}.

\subsection{Collaborative Meta-KG Construction}\label{sec:cmkg}

Directly enhancing the recommender system by the original collaborative knowledge graph $\mathcal{G}$ has two drawbacks. 1) The edges between items and entities are still sparse. Message-passing on the sparse graph is neither efficient nor effective. 
2) The learning of entity embedding burdens the training procedure. To tackle the two drawbacks, we propose to construct Collaborative Meta-KG (CMKG) $\mathcal{G}_{cmkg}$ from the original collaborative knowledge graph $\mathcal{G}$, which is defined as:
\begin{definition}
\textbf{Collaborative Meta-KG,} $\mathcal{G}_{cmkg}=\{(u,i)|u\in \mathcal{U}, i\in \mathcal{I}\}\cup\{(i,i*)|i,i* \in \mathcal{I}\}$, where $(i,i*)$ indicates the meta-edges between items built from knowledge graph. Each $\mathcal{G}_{cmkg}$ contains the historical user-item interactions and the specific item similarity extracted from the knowledge graph.
\end{definition}

\begin{figure}
     \includegraphics[width=0.5\textwidth]{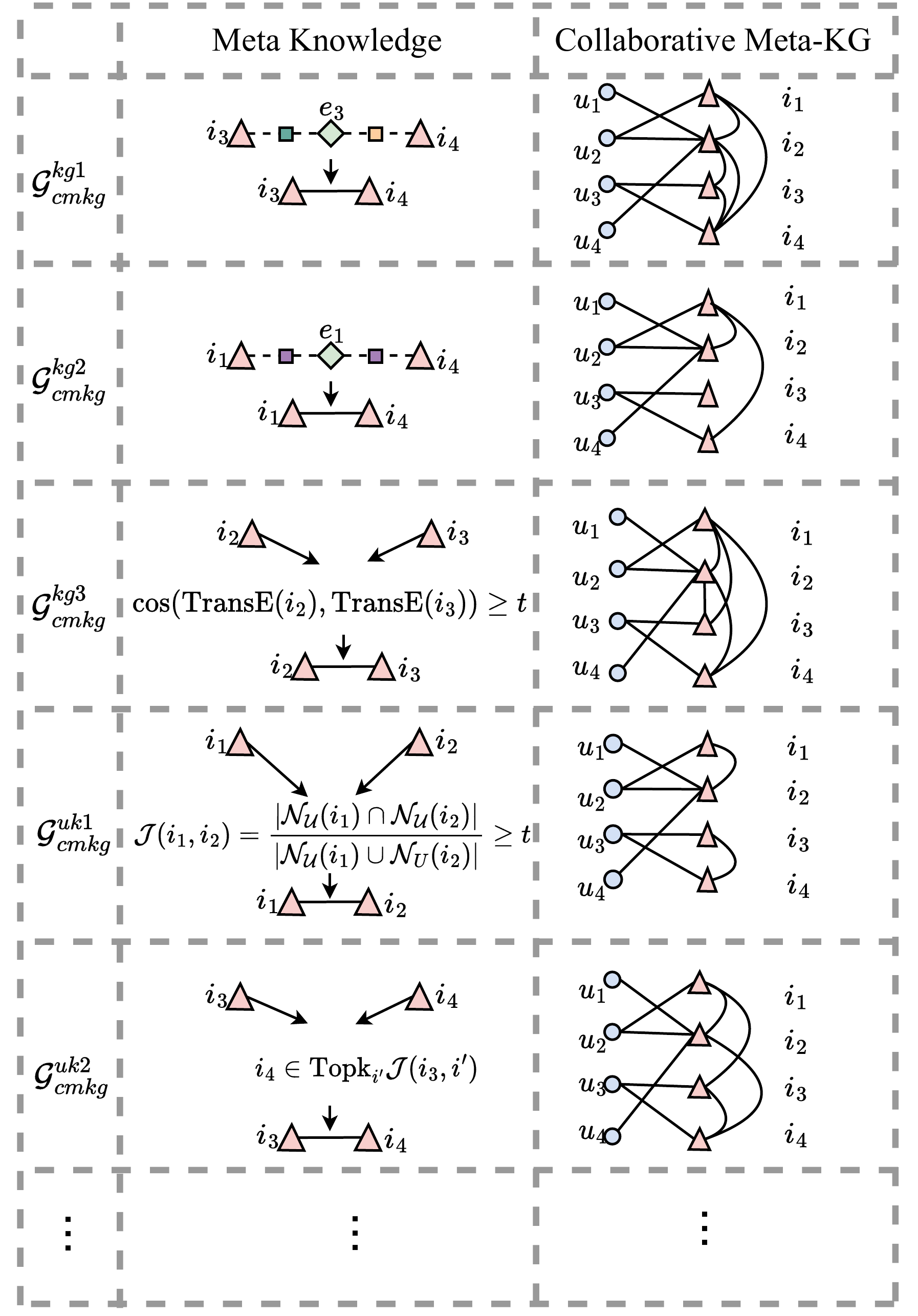}
     \caption{Collaborative Meta-KG Construction.}
     \label{fig:CMKG}
\end{figure}

$\mathcal{G}_{cmkg}$ is constructed from $\mathcal{G}$. Based on explicit meta-knowledge, we encode the knowledge graph as multiple $\mathcal{G}_{cmkg}$. In this way, we drop the entities in KG and encode the KG information as edges directly between items in $\mathcal{G}_{cmkg}$. This leads to denser interactions and enables a direct message-passing between items. As shown in Fig.~\ref{fig:CMKG}, we generate $3$ $\mathcal{G}_{cmkg}$ from the knowledge graph:
\begin{itemize}
    \item $\mathcal{G}_{cmkg}^{kg1}$: Two items are similar when connected to the same entity. E.g., we transform $(i_3\stackrel{R_1}{--}e_3\stackrel{R_2}{--}i_4)$ relation in $\mathcal{G}$ to a direct edge $(i_3--i_4)$ in $\mathcal{G}_{cmkg}^{kg1}$.
    \item $\mathcal{G}_{cmkg}^{kg2}$: Two items are more similar if they are connected to the same entity under the same relation. E.g., we transform $(i_1\stackrel{R_1}{--}e_1\stackrel{R_1}{--}i_4)$ relation in $\mathcal{G}$ to a direct edge $(i_1--i_4)$ in $\mathcal{G}_{cmkg}^{kg2}$.
    \item $\mathcal{G}_{cmkg}^{kg3}$: Items are similar if they share similar semantic meanings in the knowledge graph. To this end, we obtain the entities embedding by training a TransE~\cite{bordes2013translating} model on the knowledge graph. Then we build edges between two items if the cosine similarity of their TransE embedding is larger than a threshold $t$. E.g., we build an edge between $i_2$ and $i_3$ because their cosine similarity on TransE embedding is larger than a threshold $t$.
\end{itemize}
We can also generate $\mathcal{G}_{cmkg}$ from the user-item interactions. \modelname generates $2$ $\mathcal{G}_{cmkg}$ based on co-purchase behavior:
\begin{itemize}
    \item $\mathcal{G}_{cmkg}^{uk1}$: Items are similar if they share similar common users. We compute the Jaccard similarity between item pairs based on interacted users. Item's user neighborhood is more similar with a higher Jaccard similarity. E.g., we construct an edge between $i_1--i_2$ in $\mathcal{G}_{cmkg}^{uk1}$ because their Jaccard similarity is larger than $t$.
    \item $\mathcal{G}_{cmkg}^{uk2}$: Based on Jaccard similarity, we construct the edge between one item and its Top K most similar items. E.g., we construct $i_3--i_4$ in $\mathcal{G}_{cmkg}^{uk2}$ because $i_4$ is in $i_3$'s Top K similar items.
\end{itemize}

All $\mathcal{G}_{cmkg}$ share the historical user-item interactions and differ from the edges between items. Different $\mathcal{G}_{cmkg}$ utilizes different information sources and Meta Knowledge. $\mathcal{G}_{cmkg}^{kg1}$, $\mathcal{G}_{cmkg}^{kg2}$, $\mathcal{G}_{cmkg}^{kg3}$ are built based on knowledge graph. $\mathcal{G}_{cmkg}^{uk1}$ and $\mathcal{G}_{cmkg}^{uk2}$ are built based on user-item interaction graph. The $\mathcal{G}_{cmkg}$ can be built based on different kinds of Meta Knowledge. In \modelname, we illustrate the building method in $5$ differently. It is flexible enough to contain more information by transforming the Meta Knowledge into item similarities.

\subsection{Light Graph Convolution Encoder}
Light graph convolution~\cite{lightgcn} (LGC) has been shown effective on the user-item bipartite graph. The direct embedding passing between users and items explicitly aggregates the collaborative filtering signal. \modelname applies LGC on each generated $\mathcal{G}_{cmkg}$ in Section~\ref{sec:cmkg}.

The embedding layer is built before the convolution to represent the users and items. The embedding layer is a look-up table that maps user/item ID to a dense vector:
\begin{equation}
    \mathbf{E}=\left(\mathbf{e}_1, \mathbf{e}_2, \ldots, \mathbf{e}_{|\mathcal{U}|+|\mathcal{I}|}\right), \\
\end{equation}
where $\mathbf{e}\in \mathbb{R}^d$ is the $d$-dimensional vector to represent a user/item. The embedding indexed from the loop-up table is before the graph convolution. Then they are fed into the light graph convolution to aggregate neighbor's information:
\begin{equation}
      \mathbf{e}_n=\sum_{v\in \mathcal{N}_n}\frac{1}{\sqrt{|\mathcal{N}_n|}\sqrt{|\mathcal{N}_v|}}\mathbf{e}_v,\\  
\end{equation}
where $n,v$ can represent both user and item in $\mathcal{G}_{cmkg}$, $\mathcal{N}_n$ is the neighbor set of node $n$. The LGC can be stacked in several layers, and we omit the notation for layers for simplicity. We perform the convolution separately for each $\mathcal{G}_{cmkg}$. After the convolution, we have $[\mathbf{e}_n^{\mathcal{G}_{cmkg}^{kg1}},\mathbf{e}_n^{\mathcal{G}_{cmkg}^{kg2}},\mathbf{e}_n^{\mathcal{G}_{cmkg}^{kg3}},\mathbf{e}_n^{\mathcal{G}_{cmkg}^{uk1}},\mathbf{e}_n^{\mathcal{G}_{cmkg}^{uk2}}]$ for node $n$.

\subsection{Channel Attention}
Each $\mathcal{G}_{cmkg}$ is one channel for model to encode KG information. Different channels encode node embedding from different $\mathcal{G}_{cmkg}$, and contain different information. The channel attention module is a readout function to aggregate the embedding learned on different $\mathcal{G}_{cmkg}$:
\begin{align}
    \mathbf{e_n} &= \text{Readout}([\mathbf{e}_n^{\mathcal{G}_{cmkg}^{kg1}},\mathbf{e}_n^{\mathcal{G}_{cmkg}^{kg2}},\mathbf{e}_n^{\mathcal{G}_{cmkg}^{kg3}},\mathbf{e}_n^{\mathcal{G}_{cmkg}^{uk1}},\mathbf{e}_n^{\mathcal{G}_{cmkg}^{uk2}}]) \\
    &= \sum_{g} a^{g}\mathbf{e}^{\mathcal{G}_{cmkg}^g},
\end{align}
where $a^{g}$ is the attention weight learned for $\mathcal{G}_{cmkg}^g$, and is calculated by:
\begin{equation}
    a^{g} = \frac{\text{exp}(\langle\mathbf{W}_{\text{Att}}, \mathbf{e}^{\mathcal{G}_{cmkg}^g}\rangle)}{\sum_{g'} \text{exp}(\langle\mathbf{W}_{\text{Att}}, \mathbf{e}^{\mathcal{G}_{cmkg}^{g'}}\rangle)}
\end{equation}
where $\mathbf{W}_{\text{Att}}\in \mathbb{R}^{d}$ is the parameter to compute attention weight. The channel attention module learns different weights for different channels to optimize the loss function. It can effectively combine information extracted from different $\mathcal{G}_{cmkg}$.

\subsection{Prediction}
The aggregated embedding for $u$ and $i$ contain information from different $\mathcal{G}_{cmkg}$, and we compute the dot product as the probability that $u$ will purchase $i$:
\begin{equation}
    \hat{y}_{u,i}=\mathbf{e}_{u}\cdot\mathbf{e}_{i}.
\end{equation}

We randomly sample one negative item over the whole item set for each positive interaction and compute the Bayesian Personalized Ranking (BPR) loss~\cite{bpr} for optimization:
\begin{equation}
    \mathcal{L}=-\sum_{(u,i)\in \mathcal{G}_{rec}, j\in \mathcal{I}\setminus\mathcal{N}_u}\log \sigma(\hat{y}_{u,i}-\hat{y}_{u,j}) + \lambda ||\Theta||^2_2,
\end{equation}
where $j$ is the randomly sampled negative item, $\lambda$ is the regularization hyper-parameter, and $\Theta$ contains all the parameters.

\subsection{Model Parameter Analysis}
\begin{table}
  \caption{Model Size Comparison}
  \label{tab:model_size}
  \centering
  \begin{tabular}{c c}
        \hline
        \textbf{Model} & \textbf{Parameters} \\
        \hline
        KGAT & $(|\mathcal{U}| + |\mathcal{I}| + |\mathcal{E}| + (1+d)|\mathcal{R}| + dl + l)d$ \\
        KGIN & $(|\mathcal{U}| + |\mathcal{I}| + |\mathcal{E}| + |\mathcal{R}|)d + |\mathcal{R}||\mathcal{P}| $\\
        \modelname & $(|\mathcal{U}| + |\mathcal{I}|)d$ \\
        \hline
  \end{tabular}
\end{table}

We compare the parameter size of \modelname with KG-enhanced recommendation models. The comparison is shown in Table~\ref{tab:model_size}. $\mathcal{U},\mathcal{I},\mathcal{E},\mathcal{R}$ represent the user, item, entity and relation set, respectively. $\mathcal{P}$ is the intent set in KGIN. $l$ is the number of layers and $d$ is the embedding size. \modelname has the fewest parameters, and it only needs the user and item embedding table. All the other KG-enhanced methods need much more parameters to represent the entities. \modelname abandons the redundant design of entity embedding and enables direct aggregation between items. As shown in Table~\ref{tab:comparison}, \modelname achieves the best performance with the fewest parameters. It testifies the \modelname is effective and efficient.

\section{Experiment}
Extensive experiments are conducted on $4$ real-world datasets to answer the following research questions (\textbf{RQs}):
\begin{itemize}
    \item \textbf{RQ1}: Is \modelname effective in knowledge graph enhanced recommendation?
    \item \textbf{RQ2}: What are the results on different Collaborative Meta-KGs?
    \item \textbf{RQ3}: Can \modelname  cope with the cold-start recommendation by considering more side information?
    \item \textbf{RQ4}: What is the influence of different experiment setting on \modelname?
\end{itemize}
\subsection{Experiment Setting}
\subsubsection{Datasets}
To test the performance of \modelname, we select four datasets with knowledge graphs as side information during the evaluation. The data statistics are shown in Table~\ref{tab:data}. The size of four datasets varies from small to large, which are introduced as follows:
\begin{itemize}
    \item Music~\cite{kgcn}: It contains musician listening information of the Last.fm online platform~\footnote{https://www.last.fm/}.
    \item Book~\cite{kgcn}: It contains the rating from users to books of the Book-Crossing platform~\footnote{https://www.bookcrossing.com/}.
    \item Amazon~\cite{amazon}: It is a widely used dataset for studying e-commerce recommendations from Amazon platform~\footnote{https://www.amazon.com/}. To ensure the data quality, $10$-core setting is adopted, \textit{i.e.}, each user/item has at least $10$ interactions.
    \item Yelp~\cite{amazon}: It is a dataset from Yelp platform~\footnote{https://www.yelp.com/}, where users can check in and rate local businesses such as restaurants and bars. We also apply $10$-core setting to ensure data quality.
\end{itemize}

All the datasets contain both user-item interactions and the related knowledge graph.


\begin{table}
\centering
  \caption{Statistics of the Datasets}
  \label{tab:data}
  \begin{tabular}{l c c c c}
        \hline
        \textbf{Dataset} & \textbf{Music} & \textbf{Book} & \textbf{Amazon} & \textbf{Yelp} \\
        \hline
        \textbf{Users} & 1,872  & 17,860  & 45,113 & 15,959 \\
        \textbf{Items} & 3,846  & 14,967  & 24,695 & 43,214 \\
        \textbf{Interactions} & 42,348 & 139,746 & 426,503 & 341,211 \\
        \textbf{Density} & 0.59\% & 0.05\% & 0.038\% & 0.049\% \\
        \hline
        \textbf{Entities} & 3,846 & 77,891 & 113,269 & 134,175  \\
        \textbf{Relations} & 60 & 25 & 39 & 42 \\
        \hline
  \end{tabular}
\end{table}

\subsubsection{Baselines}\label{sec:baselines}
To make a thorough comparison, different kinds of methods are selected as baselines:

\begin{itemize}
    \item FM~\cite{fm}: It is a supervised learning algorithm based on linear regression and matrix factorization.
    \item FM-KG~\cite{kgat}: It adds the knowledge graph entity connection vector to enrich the item feature before the factorization machine.
    \item GCN~\cite{gcn}: It is the most widely used graph neural network. It makes a direct GCN convolution on the user-item bipartite graph to obtain node embedding.
    \item SGConv~\cite{sgconv}: It simplifies the convolution of multiple GCN layers by multiplying the graph Laplacian matrix.
    \item LightGCN~\cite{lightgcn}: It is the state-of-the-art recommender method. By removing the linear transform and non-linear activation, LightGCN achieves faster training and higher accuracy.
    \item KGAT~\cite{kgat}: It is a graph neural network-based method that utilizes the knowledge graph as edges between items and entities. It trains directly on the knowledge in an end-to-end fashion.
    \item KGIN~\cite{kgin}: It models the intent behind user-item interactions by combining different relations in the knowledge graph. It also designs a new information aggregation scheme to integrate the related sequences of long-range connectivity.
\end{itemize}

FM and FM-KG are two traditional methods. GCN, SGConv, and LightGCN are GNN-based methods. KGAT and KGIN are two knowledge graph-enhanced methods.

\subsubsection{Metrics}
To measure \modelname's performance, Recall and NDCG are adopted as the metrics. Recall measures the fraction of retrieved users' truly interested items, and NDCG measures the model's ranking quality of the retrieved items.
For each metric, we compare the results on top-10 and top-20 recommendations.

\begin{table*}[htbp]
    \caption{Overall comparison, the best and second-best results are in bold and underlined, respectively}
    \label{tab:comparison}
    \centering
    \begin{tabular}{l|l|cc|ccc|cc|cc}
         \hline
         Dataset & Metric & FM & FM-KG & GCN & SGConv & LightGCN & KGAT & KGIN & \modelname & Improv. \\
         \hline
         \multirow{4}{*}{Music}& R@10 & 0.04805 & 0.11771 & 0.15700 & 0.17326 & \underline{0.18966} & 0.17248 & 0.16244 & \textbf{0.20809} & 9.72\% \\
         
         & R@20 & 0.09051 & 0.18290 & 0.22449 & 0.24911 & \underline{0.27453} & 0.25043 & 0.26026 & \textbf{0.29053} & 5.83\% \\
         
         & N@10 & 0.03658 & 0.07741 & 0.10864 & 0.12299 & \underline{0.13635} & 0.11557 & 0.10729 & \textbf{0.14698} & 7.80\% \\
         
         & N@20 & 0.05197 & 0.09916 & 0.13042 & 0.14768 & \underline{0.16422} & 0.14220 & 0.13856 & \textbf{0.17433} & 6.16\% \\
         \hline
         \multirow{4}{*}{Book}& R@10 & 0.05522 & 0.06017 & 0.05914 & 0.06146 & \underline{0.06302} & 0.04329 & 0.04394 & \textbf{0.07137} & 2.63\% \\
         
         & R@20 & 0.06531 & 0.07341 & 0.07775 & 0.07971 & \underline{0.09120} & 0.05166 & 0.06033 & \textbf{0.09987} & 9.51\% \\
         
         & N@10 & 0.04274 & 0.04993 & 0.05858 & 0.04539 & \underline{0.06518} & 0.04002 & 0.04146 & \textbf{0.06842} & 4.97\% \\
         
         & N@20 & 0.04606 & 0.05453 & 0.06553 & 0.05343 & \underline{0.07623} & 0.04396 & 0.04735 & \textbf{0.07962} & 4.45\% \\
         \hline

         \multirow{4}{*}{Amazon}& R@10 & 0.00760 & 0.03540 & 0.04393 & \underline{0.04907} & 0.04379 & 0.03459 & 0.04581 & \textbf{0.05106} & 4.06\% \\
         
         & R@20 & 0.01027 & 0.06030 & 0.07176 & \underline{0.07894} & 0.06804 & 0.06129 & 0.07306 & \textbf{0.08618} & 9.17\% \\
         
         & N@10 & 0.00582 & 0.02200 & 0.02883 & \underline{0.03420} & 0.02867 & 0.02252 & 0.02950 & \textbf{0.03466} & 1.35\% \\
         
         & N@20 & 0.00812 & 0.03070 & 0.03844 & \underline{0.04470} & 0.03717  & 0.03149 & 0.03908 & \textbf{0.04645} & 3.91\% \\
         \hline

         \multirow{4}{*}{Yelp}& R@10 & 0.00400 & 0.01067 & 0.02468 & \underline{0.02524} & 0.02465 & 0.02402 & 0.02034 & \textbf{0.02830} & 12.12\% \\
         
         & R@20 & 0.00668 & 0.01777 & 0.04339 & 0.04291 & \underline{0.04510} & 0.03749 & 0.03687 & \textbf{0.04884} & 8.29\% \\
         
         & N@10 & 0.00488 & 0.00945 & 0.02619 & 0.02596 & \underline{0.02621} & 0.02247 & 0.01994 & \textbf{0.02839} & 8.32\% \\
         
         & N@20 & 0.0063 & 0.01322 & 0.03579 & 0.03552 & \underline{0.03676} & 0.03001 & 0.02872 & \textbf{0.03880} & 8.41\% \\
         \hline

    \end{tabular}
\end{table*}

\subsubsection{Hyper-parameter Setting}
We randomly split the datasets into training ($80\%$), validation ($10\%$), and test ($10\%$). We tune hyper-parameters on the validation set and report experimental results on the test set. Adam is used as the optimizer. For all methods, we tune the learning rate and weight-decay within \{0.1, 0.05, 0.01, 0.005, 0.001\}, \{1e-1, 1e-2, 1e-3, 1e-4, 1e-5, 1e-6\}, respectively. For a fair comparison, we fix the embedding size as $4$ and the negative sample rate as $1$ for all methods. To avoid overfitting, we early stop the training if the model's performance on the validation set does not improve in $10$ epochs.

\begin{table}[htbp]
  \caption{Experiment Results of different Collaborative Meta-KGs (Music)}
  \label{tab:music cmkg}
  \centering
  \begin{tabular}{c c c c c}
        \hline
        \textbf{ID} & \textbf{R@10} & \textbf{R@20} & \textbf{N@10} & \textbf{N@20}\\
        \hline
        \textbf{$\mathcal{G}_{cmkg}^{kg1}$} & 0.17464 & 0.25524 & 0.12244 & 0.14927 \\
        \textbf{$\mathcal{G}_{cmkg}^{kg2}$} & 0.17831 & 0.25015 & 0.12228 & 0.14626 \\
        \textbf{$\mathcal{G}_{cmkg}^{kg3}$} & \underline{0.18958} & \underline{0.27885} & \underline{0.13087} & \underline{0.16054} \\
        \textbf{$\mathcal{G}_{cmkg}^{uk1}$} & 0.17910 & 0.24578 & 0.12619 & 0.14930 \\
        \textbf{$\mathcal{G}_{cmkg}^{uk2}$} & 0.16501 & 0.23598 & 0.11773 & 0.14127 \\
        \textbf{\modelname} & \textbf{0.20810} & \textbf{0.29053} & \textbf{0.14699} & \textbf{0.17433} \\
        \textbf{Improv.} & 9.77\% & 4.19\% & 12.32\% & 8.59\% \\
        \hline
  \end{tabular}
\end{table}

\begin{table}[htbp]
  \caption{Experiment Results of different Collaborative Meta-KGs (Book)}
  \label{tab:book cmkg}
  \centering
  \begin{tabular}{c c c c c}
        \hline
        \textbf{ID} & \textbf{R@10} & \textbf{R@20} & \textbf{N@10} & \textbf{N@20}\\
        \hline
        \textbf{$\mathcal{G}_{cmkg}^{kg1}$} & 0.05356 & 0.06893 & 0.05493 & 0.06167 \\
        \textbf{$\mathcal{G}_{cmkg}^{kg2}$} & 0.05356 & 0.06893 & 0.05494 & 0.06167 \\
        \textbf{$\mathcal{G}_{cmkg}^{kg3}$} & \underline{0.05917} & \underline{0.08145} & \underline{0.06364} & \underline{0.07209} \\
        \textbf{$\mathcal{G}_{cmkg}^{uk1}$} & 0.05356 & 0.06888 & 0.05491 & 0.06157 \\
        \textbf{$\mathcal{G}_{cmkg}^{uk2}$} & 0.05432 & 0.06889 & 0.05134 & 0.05893 \\
        \textbf{\modelname} & \textbf{0.07138} & \textbf{0.09988} & \textbf{0.06843} & \textbf{0.07962} \\
        \textbf{Improv.} & 20.64\% & 22.63\% & 7.53\% & 10.44\% \\
        \hline
  \end{tabular}
\end{table}

\begin{table}[htbp]
  \caption{Experiment Results of different Collaborative Meta-KGs (Amazon)}
  \label{tab:amazon cmkg}
  \centering
  \begin{tabular}{c c c c c}
        \hline
        \textbf{ID} & \textbf{R@10} & \textbf{R@20} & \textbf{N@10} & \textbf{N@20}\\
        \hline
        \textbf{$\mathcal{G}_{cmkg}^{kg1}$} & 0.03940 & 0.06911 & 0.02637 & 0.03650 \\
        \textbf{$\mathcal{G}_{cmkg}^{kg2}$} & \underline{0.04184} & \underline{0.06973} & 0.02699 & 0.03662 \\
        \textbf{$\mathcal{G}_{cmkg}^{kg3}$} & 0.03762 & 0.06676 & 0.02535 & 0.03522 \\
        \textbf{$\mathcal{G}_{cmkg}^{uk1}$} & 0.03739 & 0.06679 & 0.02463 & 0.03438 \\
        \textbf{$\mathcal{G}_{cmkg}^{uk2}$} & 0.04050 & 0.06883 & \underline{0.02798} & \underline{0.03762} \\
        \textbf{\modelname} & \textbf{0.05106} & \textbf{0.08619} & \textbf{0.03467} & 0.\textbf{04646} \\
        \textbf{Improv.} & 22.04\% & 23.61\% & 23.91\% & 23.50\% \\
        \hline
  \end{tabular}
\end{table}

\begin{table}[htbp]
  \caption{Experiment Results of different Collaborative Meta-KGs (Yelp)}
  \label{tab:yelp cmkg}
  \centering
  \begin{tabular}{c c c c c}
        \hline
        \textbf{ID} & \textbf{R@10} & \textbf{R@20} & \textbf{N@10} & \textbf{N@20}\\
        \hline
        \textbf{$\mathcal{G}_{cmkg}^{kg1}$} & 0.02280 & 0.04130 & 0.02366 & 0.03314 \\
        \textbf{$\mathcal{G}_{cmkg}^{kg2}$} & 0.02672 & 0.04369 & 0.02654 & 0.03546 \\
        \textbf{$\mathcal{G}_{cmkg}^{kg3}$} & 0.02703 & 0.04316 & \underline{0.02790} & \underline{0.03665} \\
        \textbf{$\mathcal{G}_{cmkg}^{uk1}$} & 0.02570 & 0.04423 & 0.02657 & 0.03603 \\
        \textbf{$\mathcal{G}_{cmkg}^{uk2}$} & \underline{0.02731} & \underline{0.04465} & 0.02663 & 0.03610 \\
        \textbf{\modelname} & \textbf{0.02831} & \textbf{0.04885} & \textbf{0.02840} & \textbf{0.03880} \\
        \textbf{Improv.} & 3.66\% & 9.41\% & 1.79\% & 5.87\% \\
        \hline
  \end{tabular}
\end{table}

\begin{figure*}[htbp]
     \centering
     \begin{subfigure}[b]{0.24\textwidth}
         \centering
         \includegraphics[width=\textwidth]{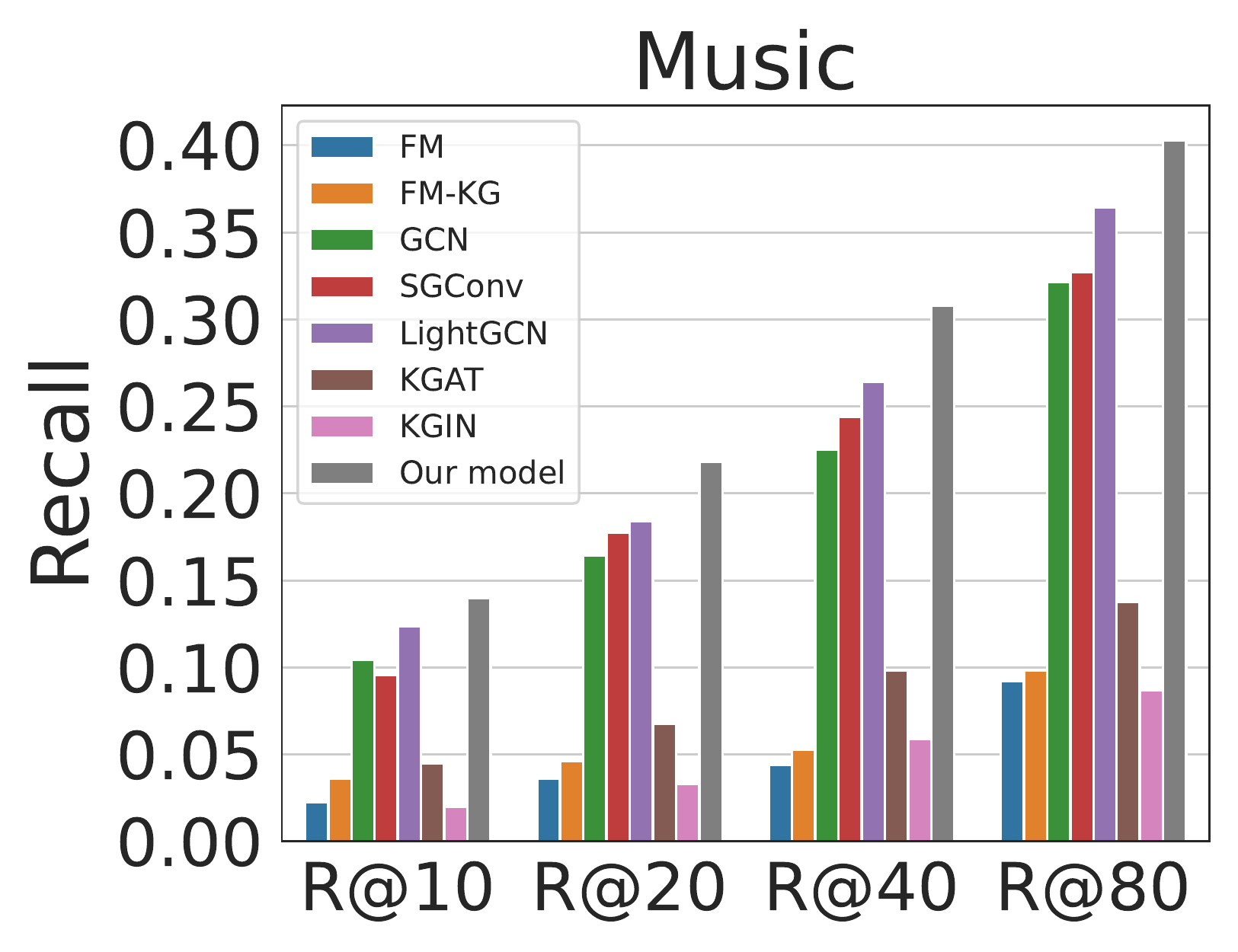}
         \label{fig:music_recall_cold}
     \end{subfigure}
     \hfill
     \begin{subfigure}[b]{0.24\textwidth}
         \centering
         \includegraphics[width=\textwidth]{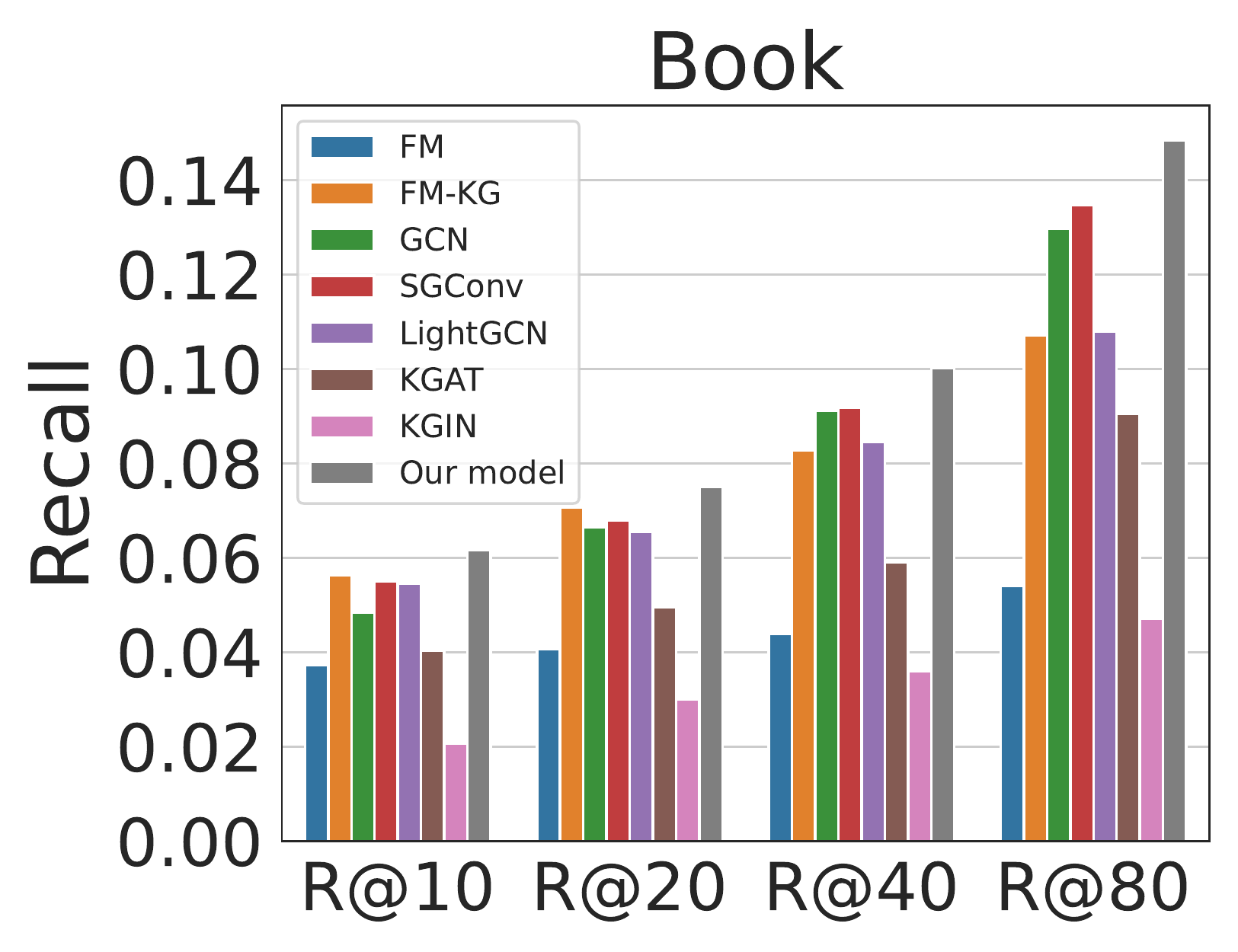}
         \label{fig:book_recall_cold}
     \end{subfigure}
     \hfill
     \begin{subfigure}[b]{0.24\textwidth}
         \centering
         \includegraphics[width=\textwidth]{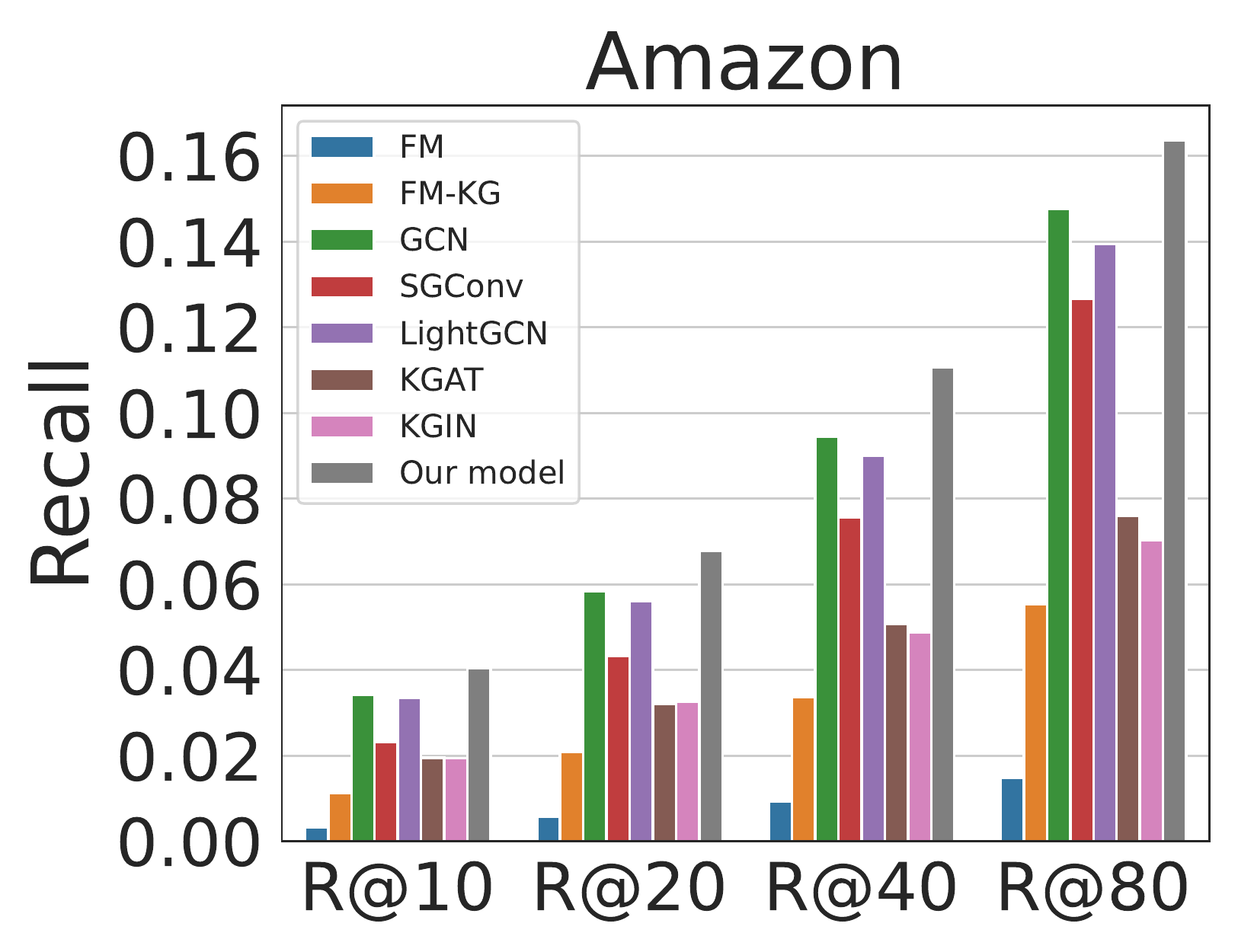}
         \label{fig:amazon_recall_cold}
     \end{subfigure}
     \begin{subfigure}[b]{0.24\textwidth}
         \centering
         \includegraphics[width=\textwidth]{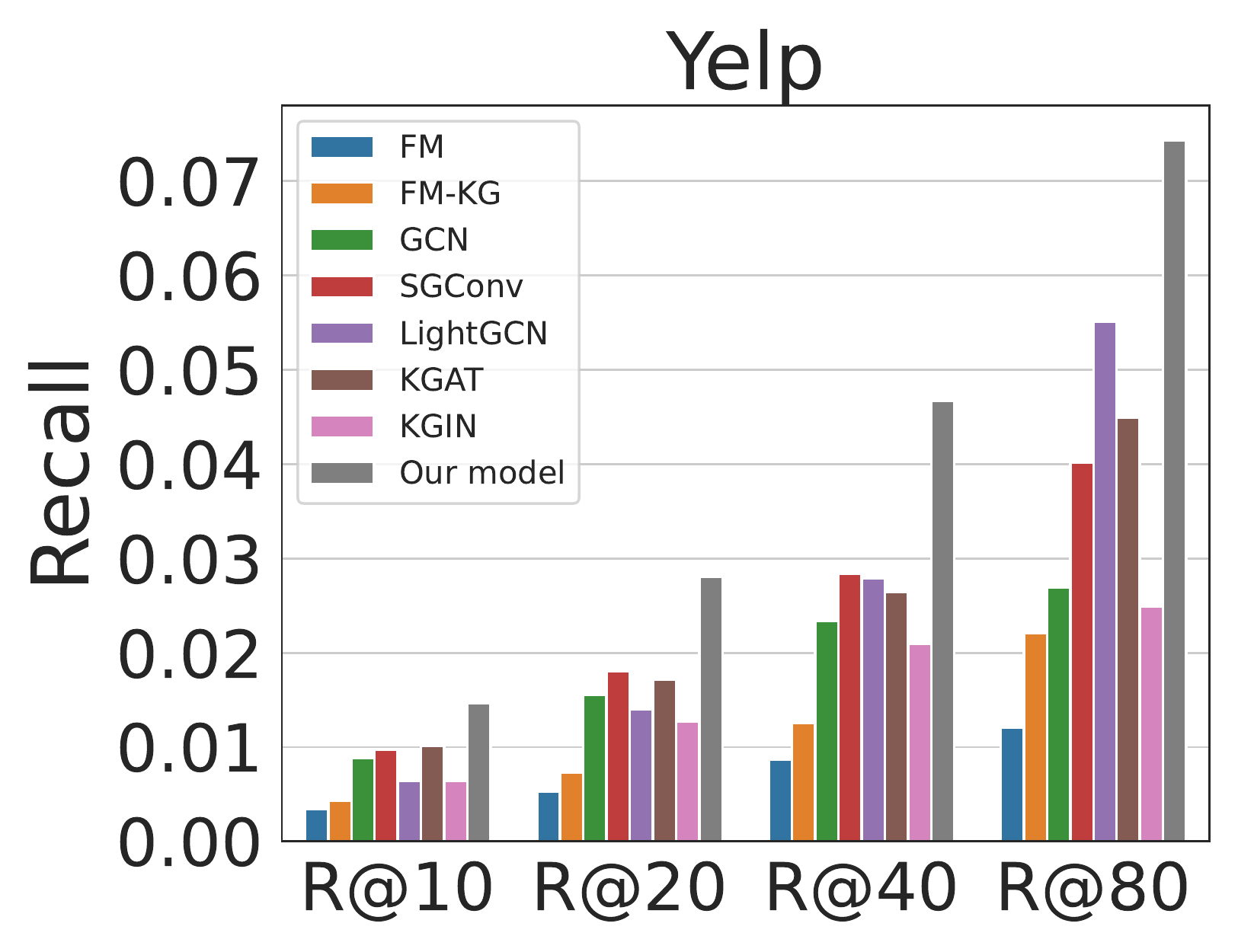}
         \label{fig:yelp_recall_cold}
     \end{subfigure}
        \caption{Recall score of different methods under cold-start setting}
        \label{fig:recall cold}
\end{figure*}

\begin{figure*}[htbp]
     \centering
     \begin{subfigure}[b]{0.24\textwidth}
         \centering
         \includegraphics[width=\textwidth]{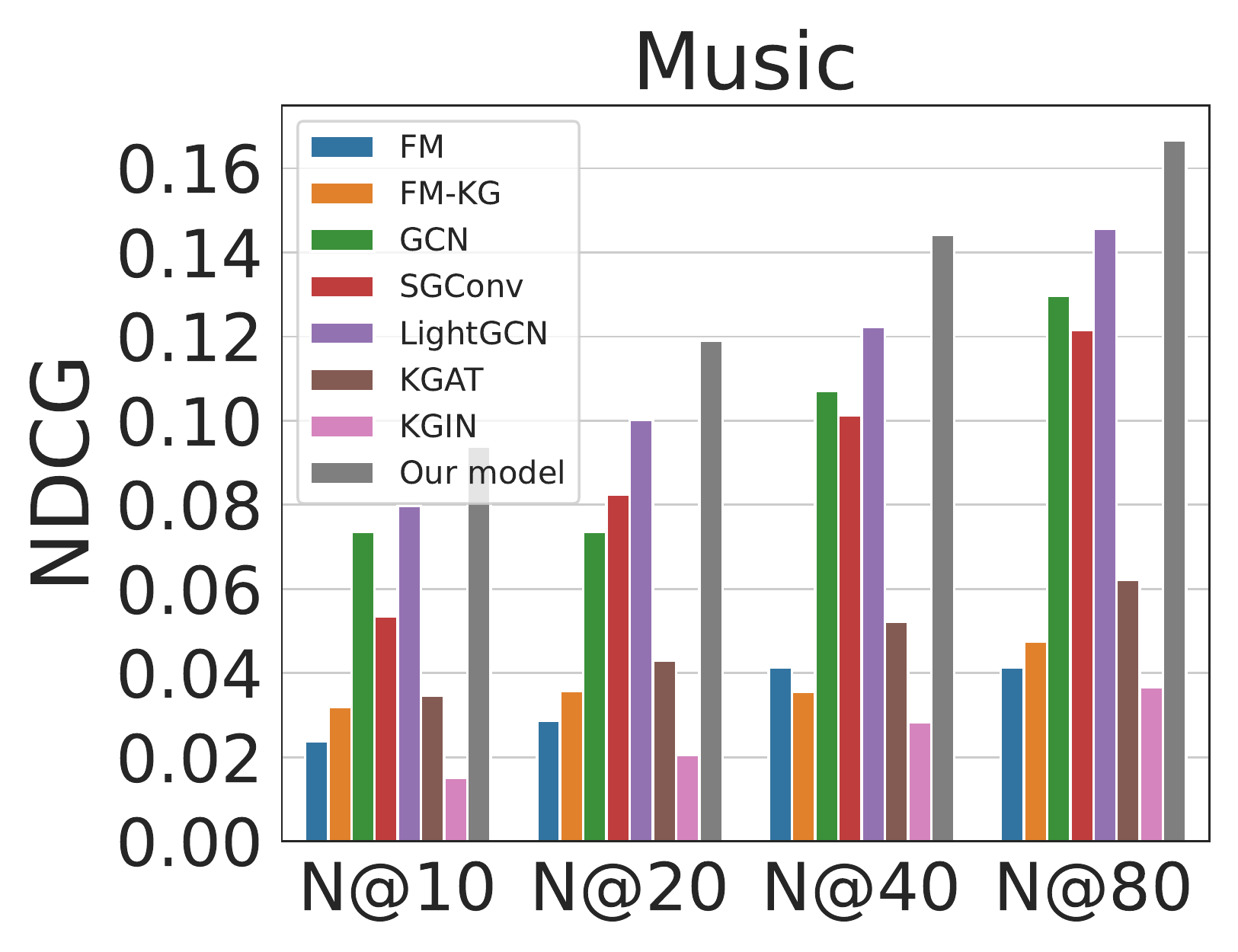}
         \label{fig:music_ndcg_cold}
     \end{subfigure}
     \hfill
     \begin{subfigure}[b]{0.24\textwidth}
         \centering
         \includegraphics[width=\textwidth]{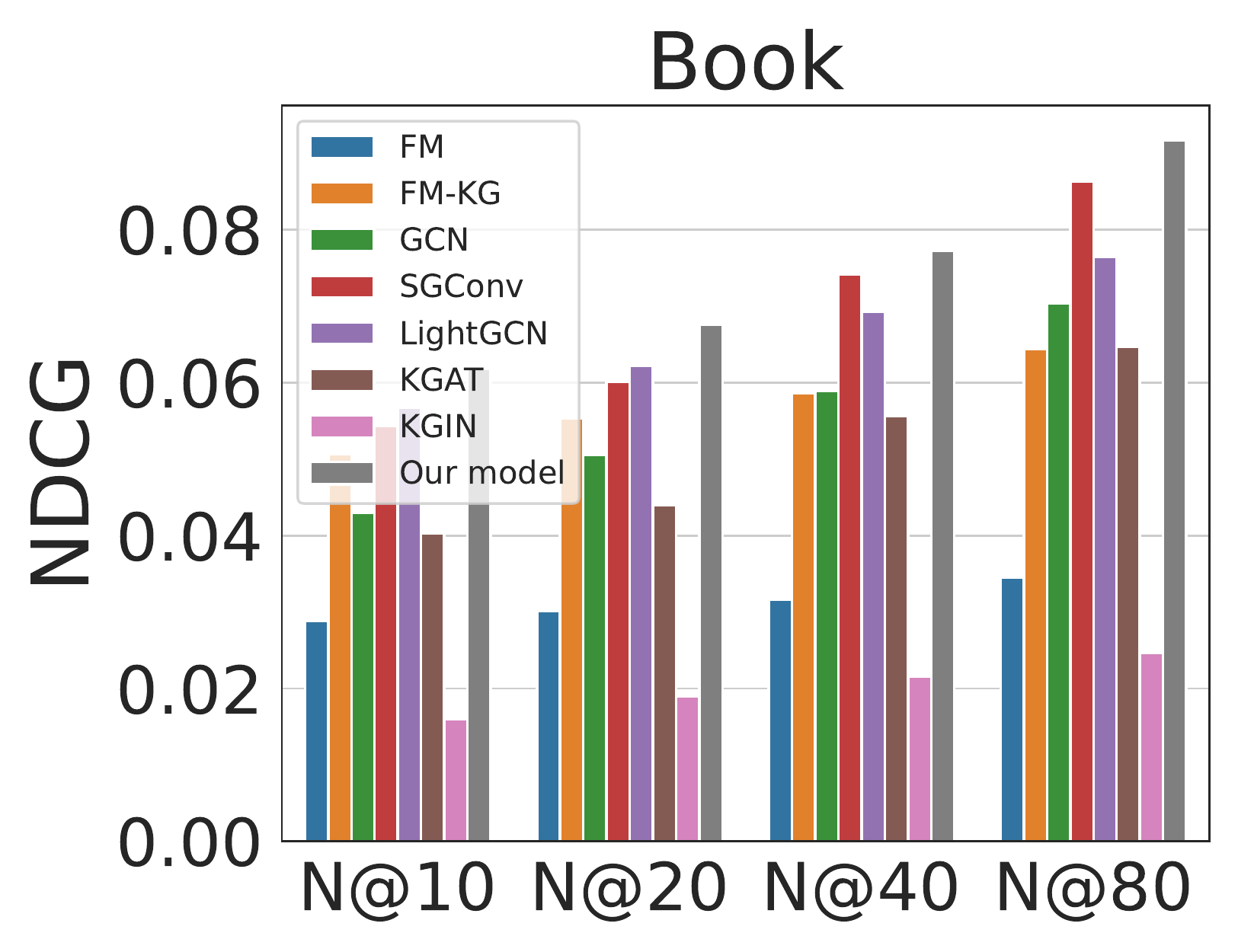}
         \label{fig:book_ndcg_cold}
     \end{subfigure}
     \hfill
     \begin{subfigure}[b]{0.24\textwidth}
         \centering
         \includegraphics[width=\textwidth]{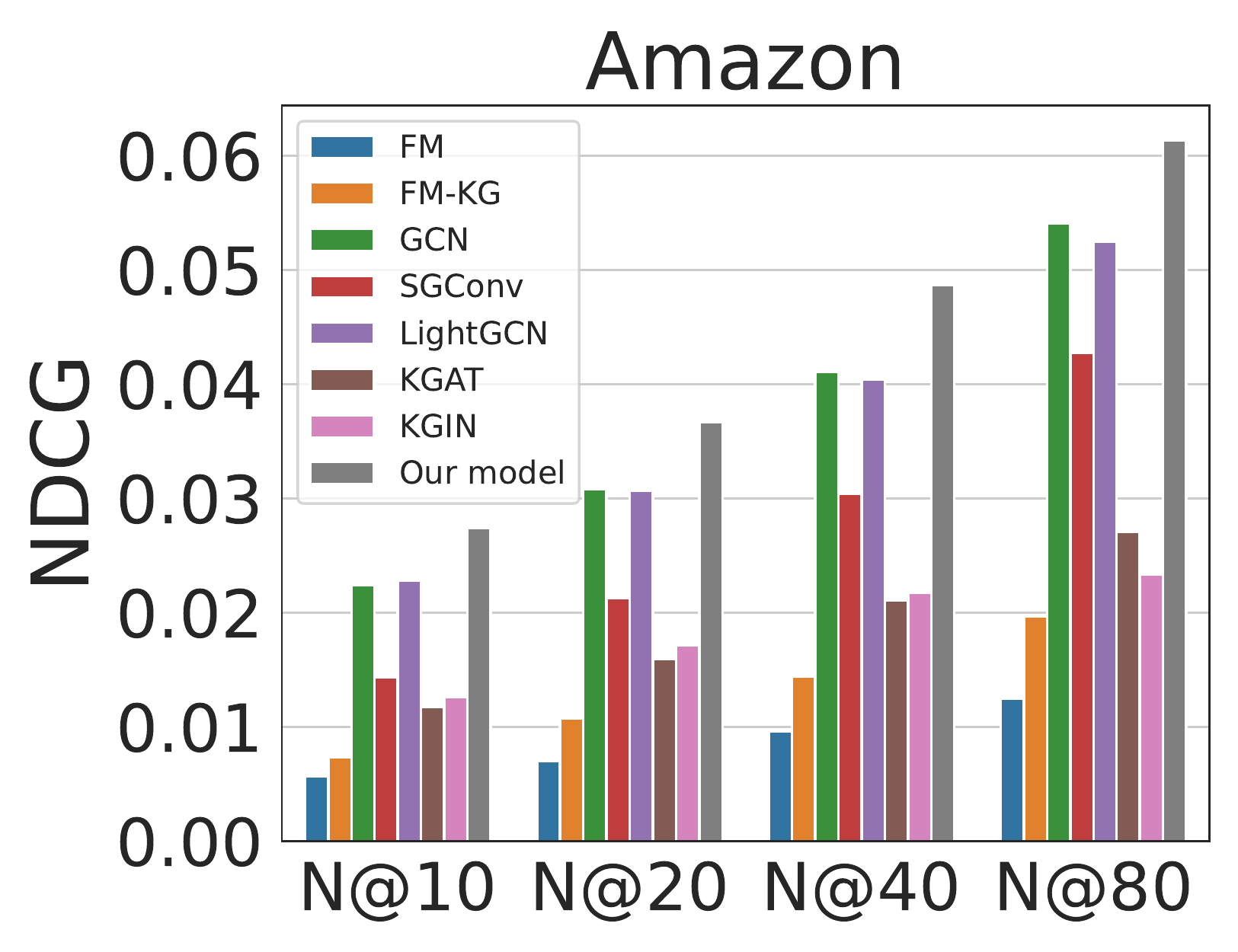}
         \label{fig:amazon_ndcg_cold}
     \end{subfigure}
     \begin{subfigure}[b]{0.24\textwidth}
         \centering
         \includegraphics[width=\textwidth]{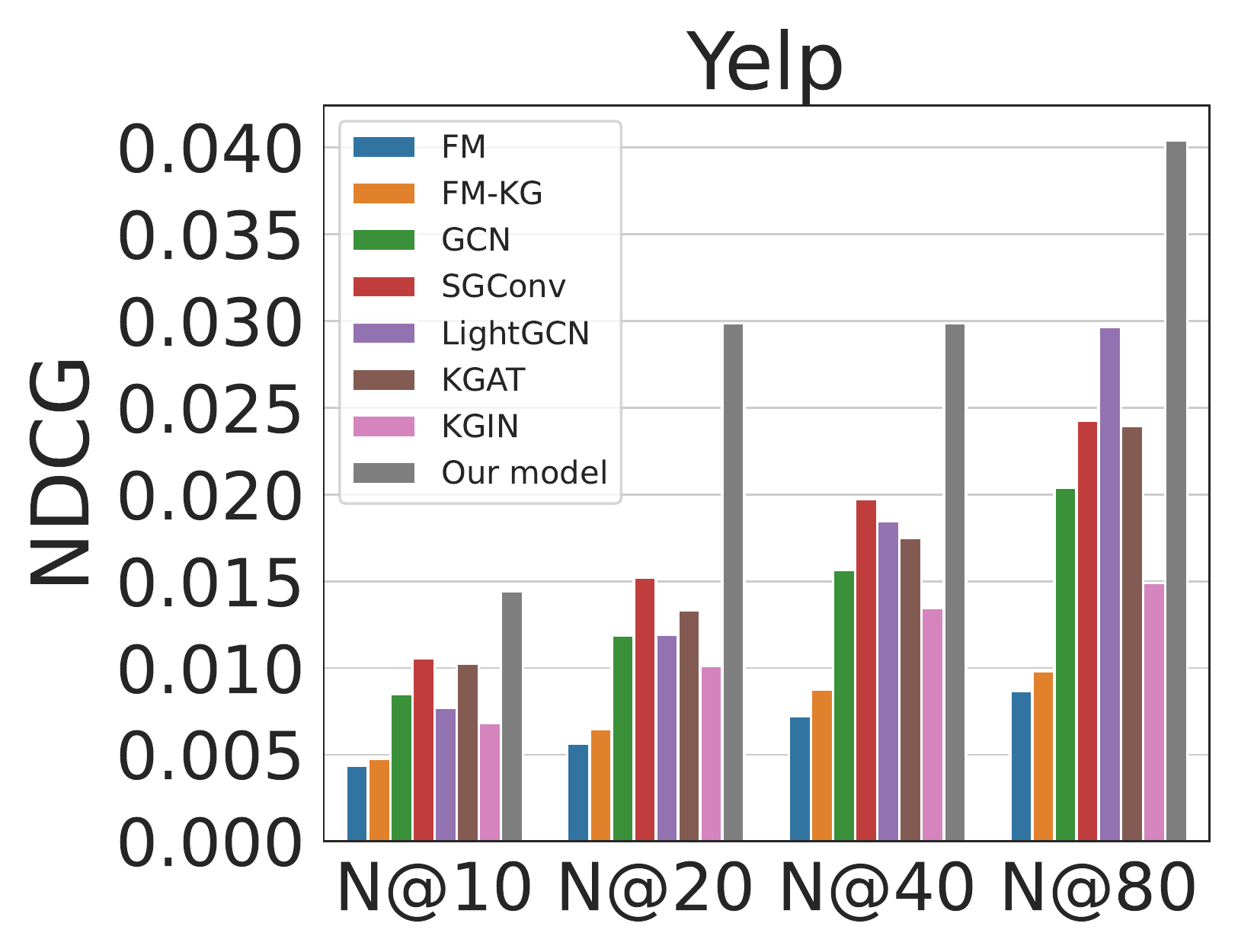}
         \label{fig:yelp_ndcg_cold}
     \end{subfigure}
        \caption{NDCG score of different methods under cold-start setting}
        \label{fig:ndcg cold}
\end{figure*}

\subsection{Overall Experiment (RQ1)}

The overall comparison of the $4$ datasets is shown in Table~\ref{tab:comparison}. We can have the following observations:
\begin{itemize}
    \item \modelname consistently outperforms the runner-up on all datasets. The improvement is over $5\%$ on all Music and Yelp datasets metrics. The huge improvement validates the effectiveness of our model.
    \item \modelname always beats LightGCN by a large margin. As \modelname and LightGCN use the same Light Graph Convolution layer, the improvement is largely brought by the KG information. It shows \modelname greatly enhances the recommender system by KG side information.
    \item KG-enhanced recommender systems (KGAT, KGIN) do not necessarily perform better than models that utilize only user-item bipartite graphs (SGConv, LightGCN). It indicates the collaborative filtering signal on the bipartite graph is most important to achieve accurate recommendations while the knowledge graph is sided information.
    \item Among all the KG enhanced methods, \modelname always achieves the best performance. It reveals that collaborative meta-kg graphs can effectively transform the knowledge graph as side information to enhance the recommender system.
\end{itemize}

\subsection{Evaluation on different Collaborative Meta-KGs (RQ2)}

We then test the model's performance on different Collaborative Meta-KGs. Experiment results on $4$ datasets are illustrated in Tables~\ref{tab:music cmkg},~\ref{tab:book cmkg},~\ref{tab:amazon cmkg},~\ref{tab:yelp cmkg} respectively. From the $4$ tables, we can have the following observations. 1) The combined model that utilizes all the Collaborative Meta-KGs always achieves the best performance. The improvement on the Amazon dataset is over $20\%$ compared with the runner-up. It shows different Collaborative Meta-KGs acquire different information and can complement each other to achieve a better performance in a joint effort.
2) The performance of each Collaborative Meta-KG does not deteriorate much compared with the joint model. It reveals the collaborative filtering signal between user-item interactions is the most important during building an effective recommender system. The knowledge graph can be utilized as side information to enhance the recommendation performance.

\subsection{Cold-start Experiment (RQ3)}
Cold-start is a severe problem in recommender systems. Without enough user-item interactions, it is difficult to obtain an informative representation. 
This experiment evaluates the model's performance under the cold-start setting, where we randomly keep only $1$ user interaction for each item in the training set. Here we select Recall and NDCG @ $\{10,20,40,80\}$ as evaluation metrics. 

Experiment results on Recall are shown in Fig.~\ref{fig:recall cold}, and results on NDCG are shown in Fig.~\ref{fig:ndcg cold}. From the results, we can observe that \modelname achieves the best performance on both Recall and NDCG over all datasets. Compared with the runner-up, the improvement of \modelname on Yelp dataset is over $30\%$. On Amazon and Music datasets, \modelname surpasses the second best over $15\%$. The lowest improvement on Book dataset is more than $10\%$. The huge improvement on all datasets validates \modelname can effectively utilize knowledge graph as side information to cope with the cold-start recommendation problem. 

We can also observe that the performance of KGAT and KGIN are even worse than the simple GCN model under the cold-start setting. It shows knowledge graph can not always provide informative information for cold-start items. The complex design of KGAT and KGIN do not contribute to better performance. With fewer parameters, \modelname achieves very huge improvement over other KG-enhanced models.

\subsection{Hyper-parameter Study (RQ4)}
In this experiment, we explore the influence of different experiment settings on \modelname. Two influential hyper-parameters are studied: 1) The number of graph convolution layers, and 2) Channel combination methods.

\subsubsection{Number of convolution layers}

\begin{figure*}[htbp]
     \centering
     \begin{subfigure}[b]{0.24\textwidth}
         \centering
         \includegraphics[width=\textwidth]{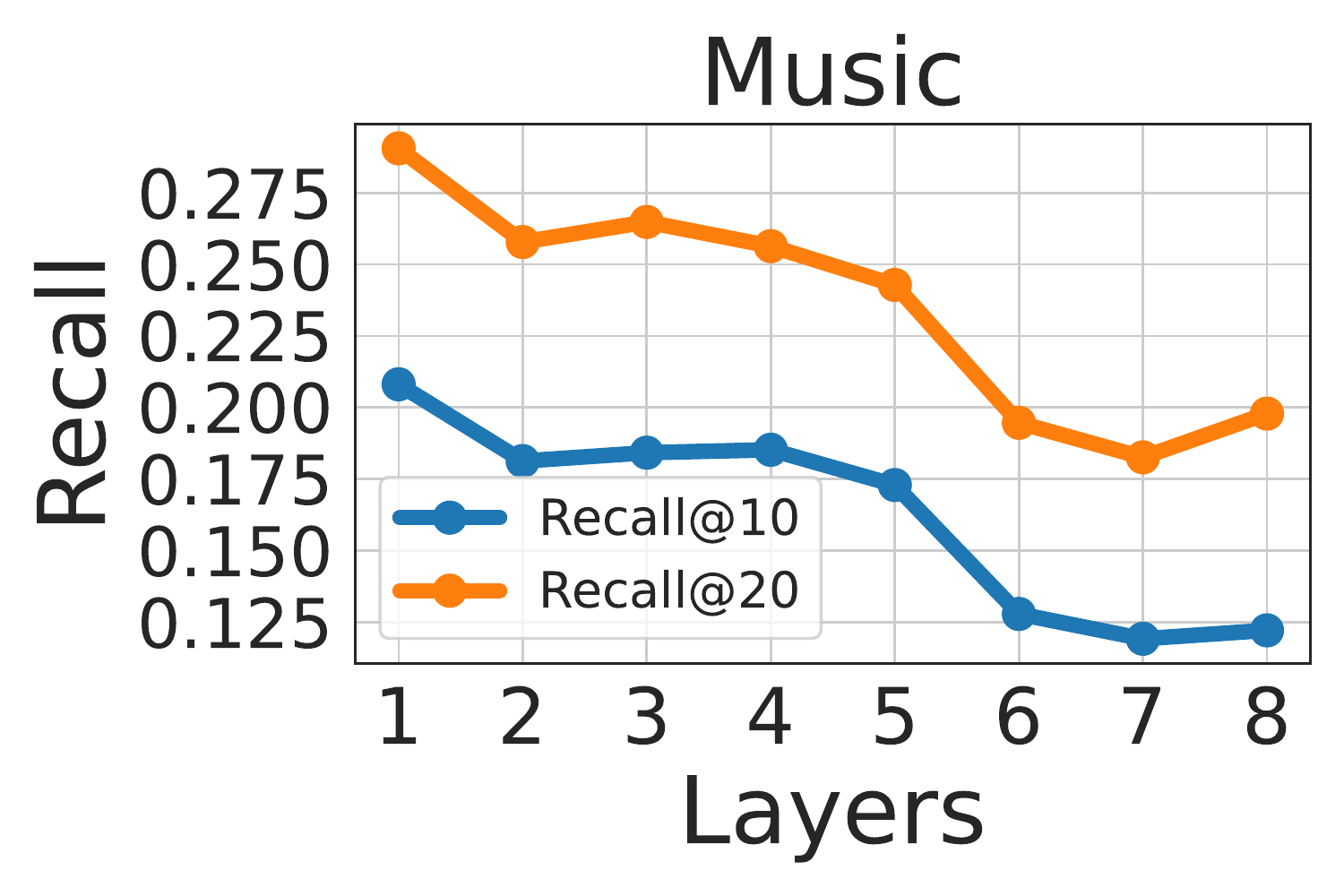}
         \label{fig:music_recall}
     \end{subfigure}
     \hfill
     \begin{subfigure}[b]{0.24\textwidth}
         \centering
         \includegraphics[width=\textwidth]{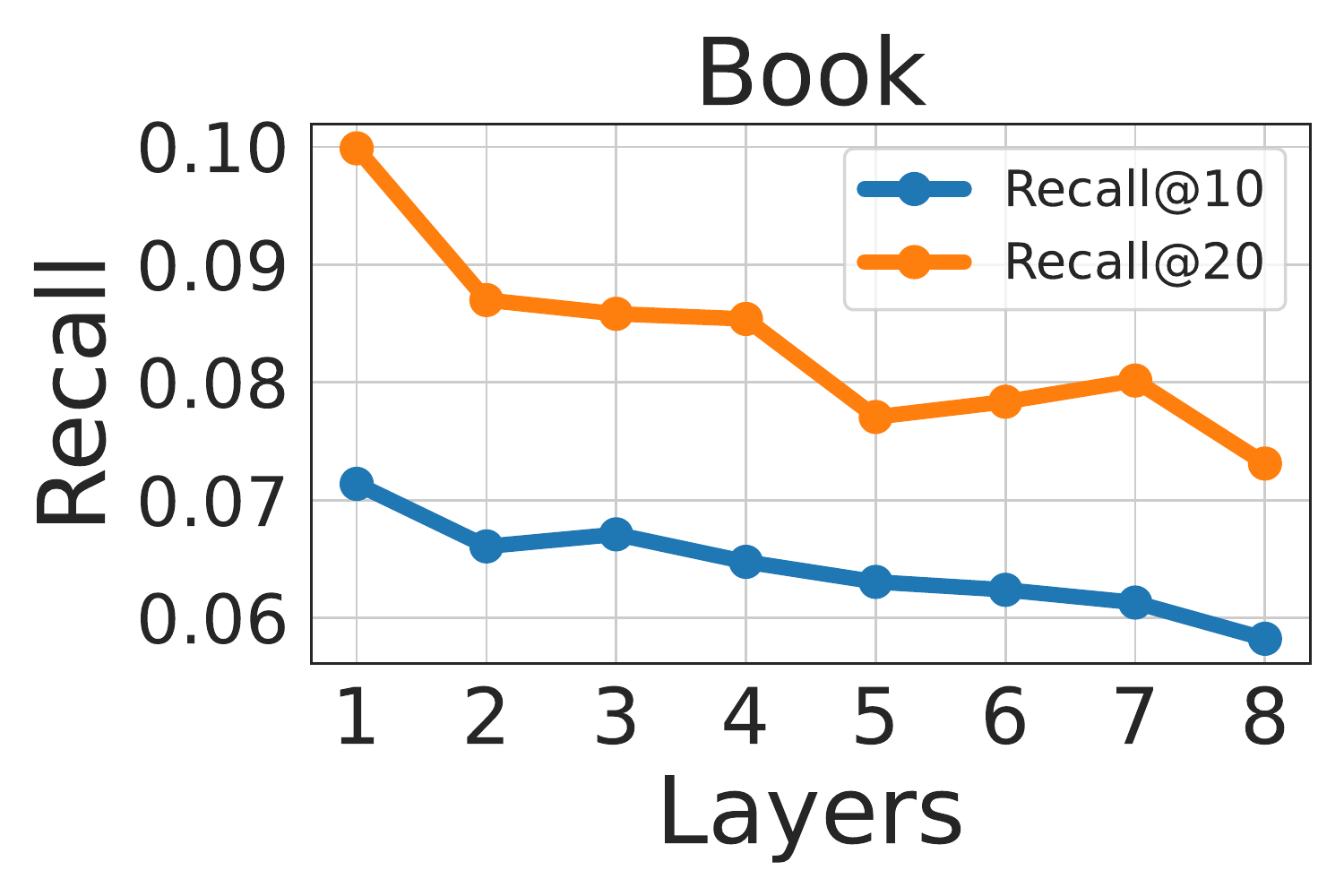}
         \label{fig:book_recall}
     \end{subfigure}
     \hfill
     \begin{subfigure}[b]{0.24\textwidth}
         \centering
         \includegraphics[width=\textwidth]{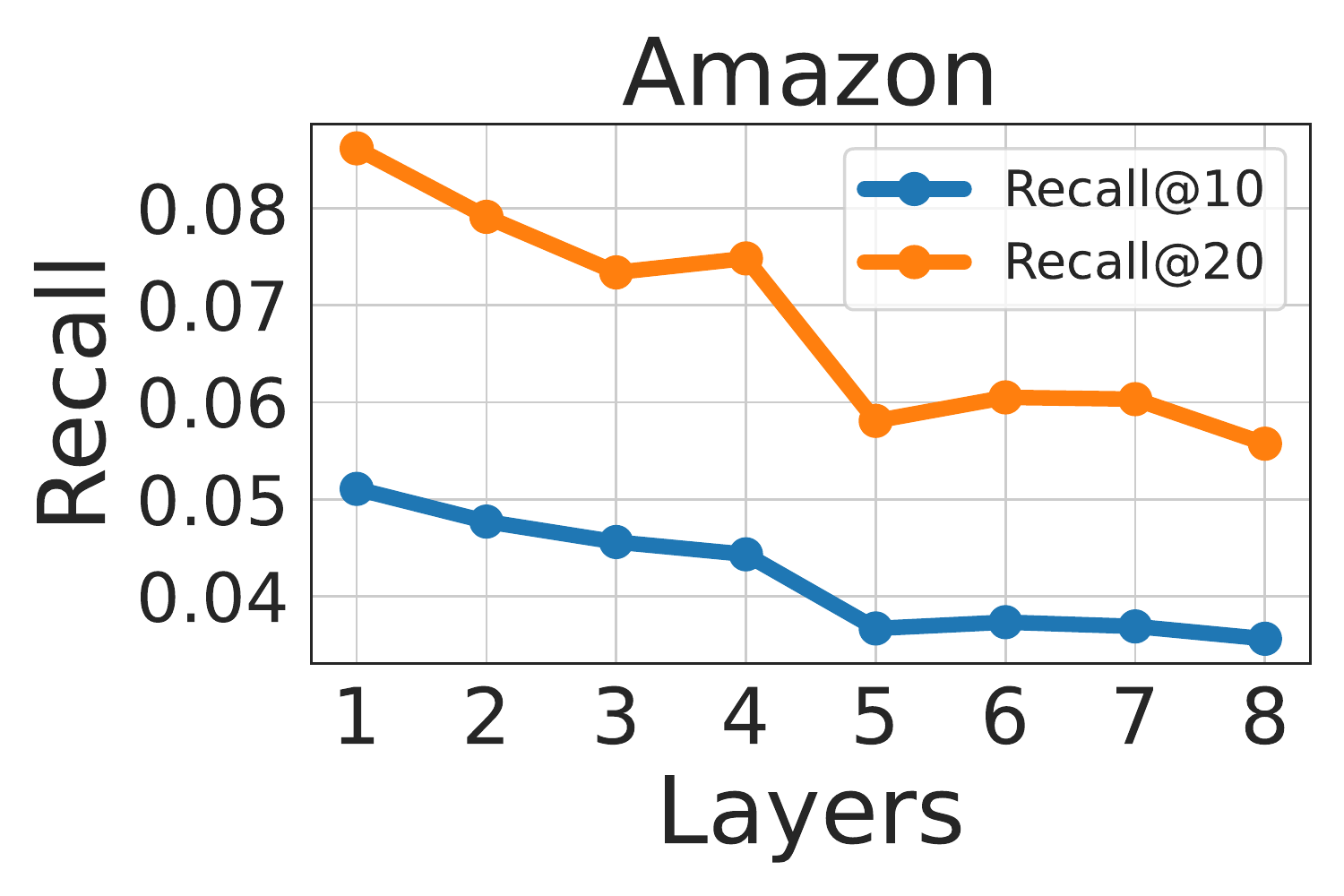}
         \label{fig:amazon_recall}
     \end{subfigure}
     \begin{subfigure}[b]{0.24\textwidth}
         \centering
         \includegraphics[width=\textwidth]{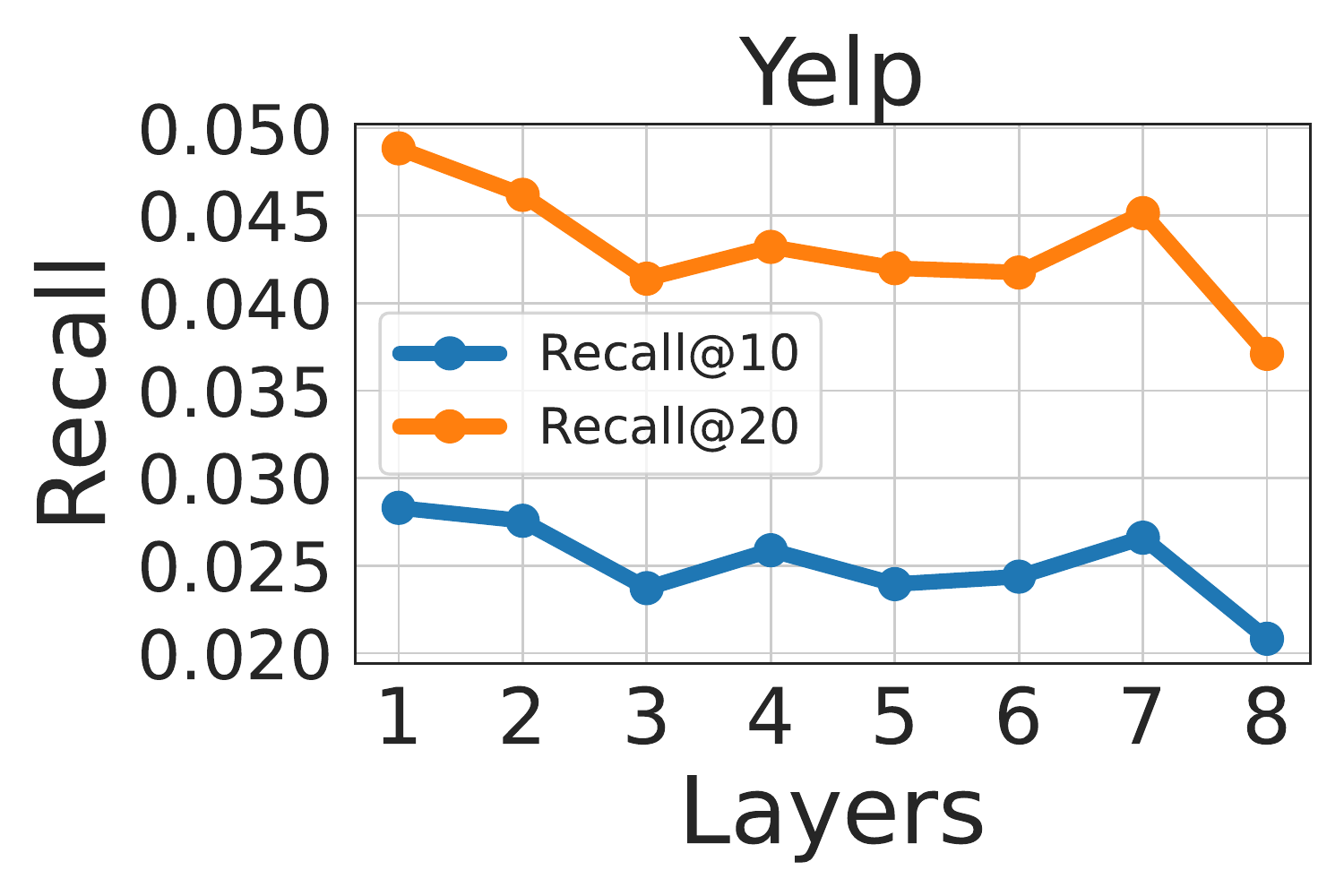}
         \label{fig:yelp_recall}
     \end{subfigure}
        \caption{Recall change with the increase of layers}
        \label{fig:recall layer}
\end{figure*}

\begin{figure*}[htbp]
     \centering
     \begin{subfigure}[b]{0.24\textwidth}
         \centering
         \includegraphics[width=\textwidth]{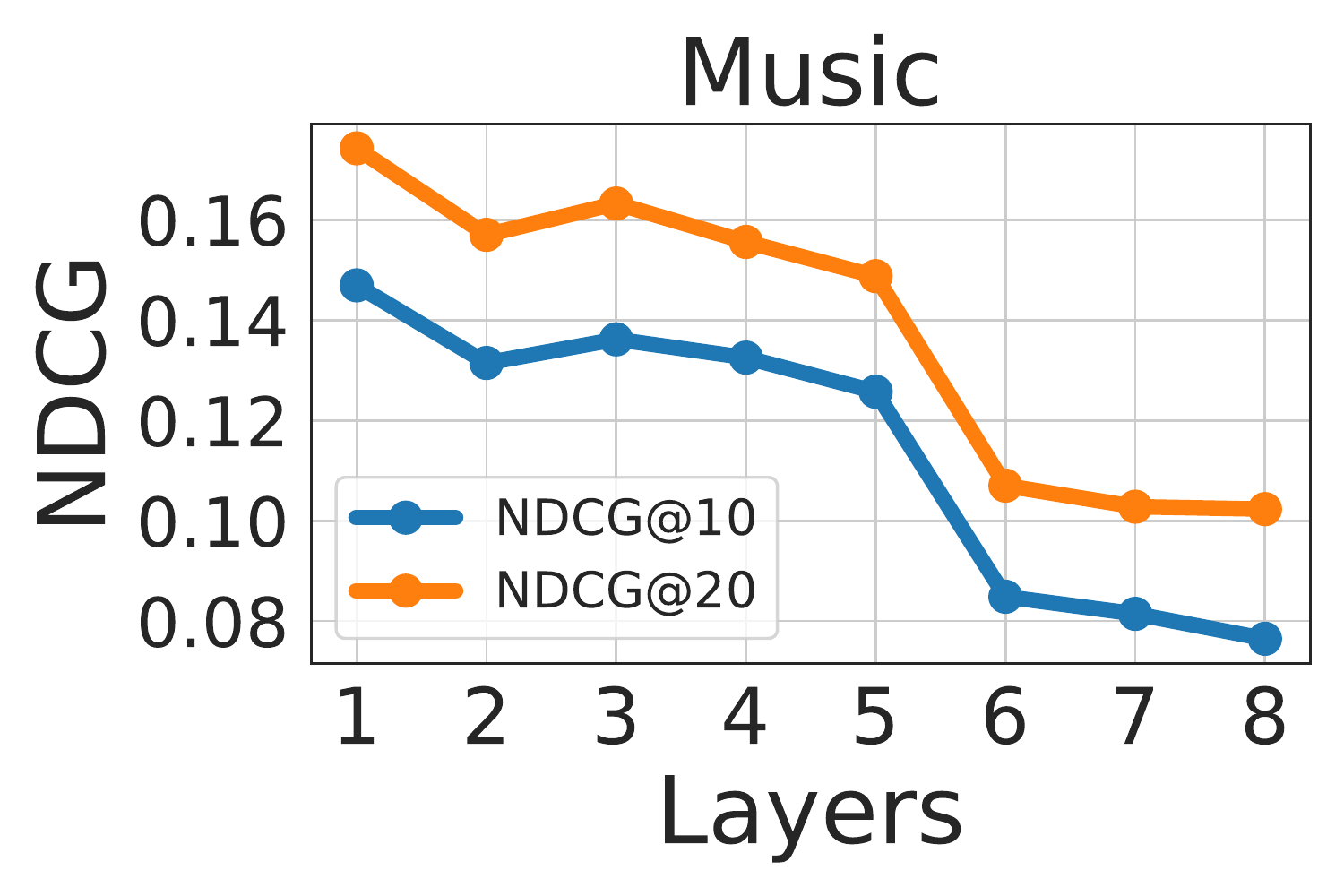}
     \end{subfigure}
     \hfill
     \begin{subfigure}[b]{0.24\textwidth}
         \centering
         \includegraphics[width=\textwidth]{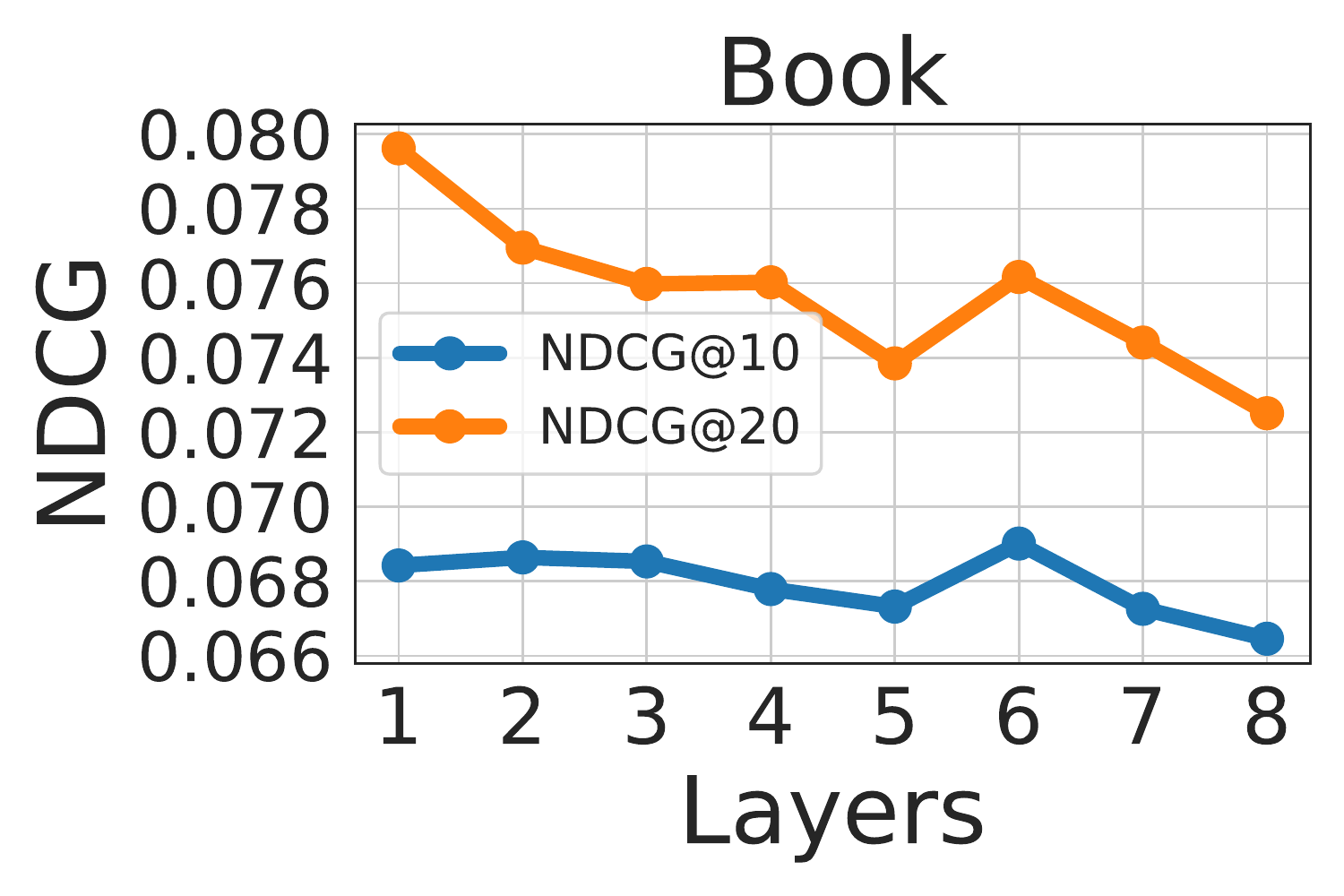}
     \end{subfigure}
     \hfill
     \begin{subfigure}[b]{0.24\textwidth}
         \centering
         \includegraphics[width=\textwidth]{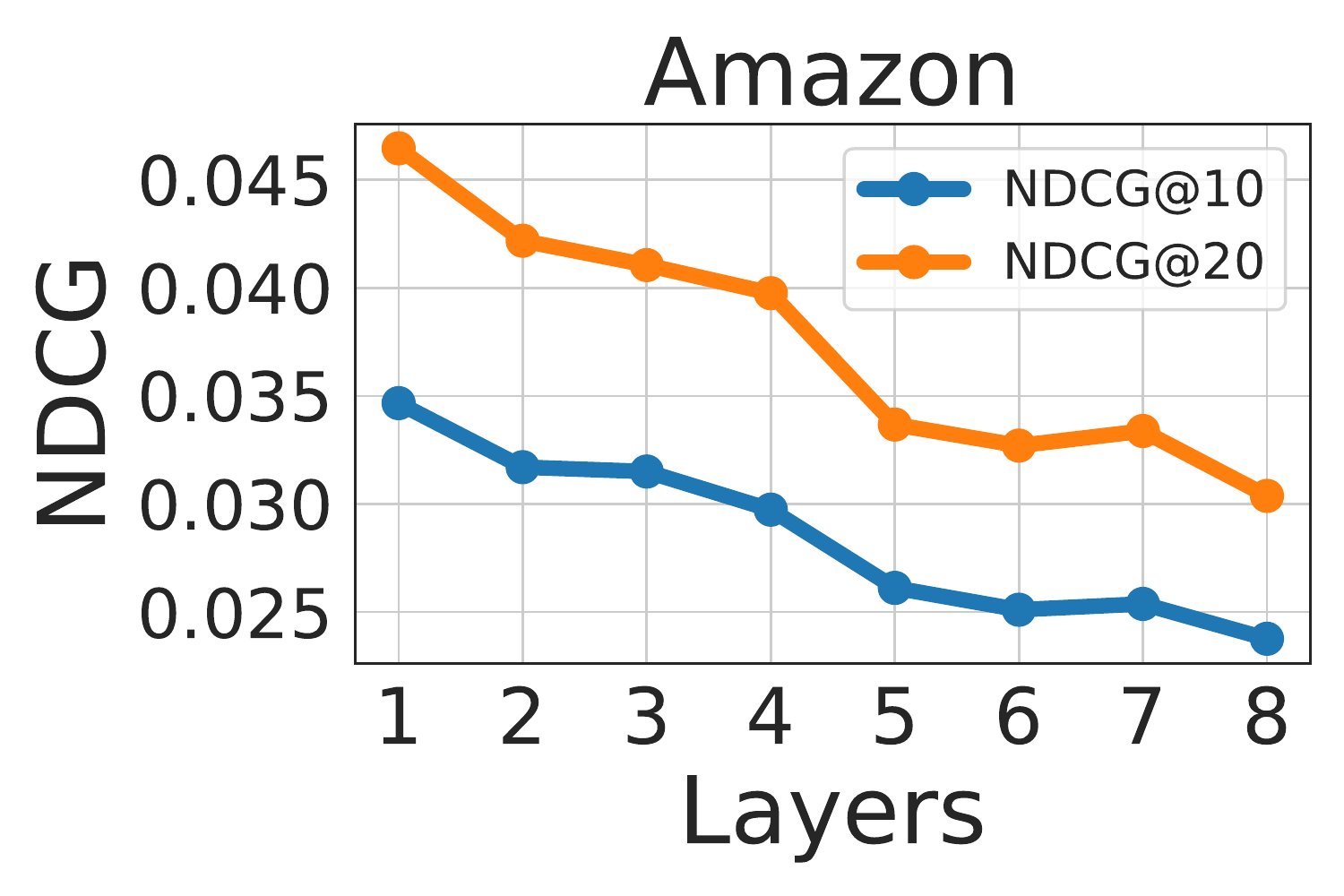}
     \end{subfigure}
     \begin{subfigure}[b]{0.24\textwidth}
         \centering
         \includegraphics[width=\textwidth]{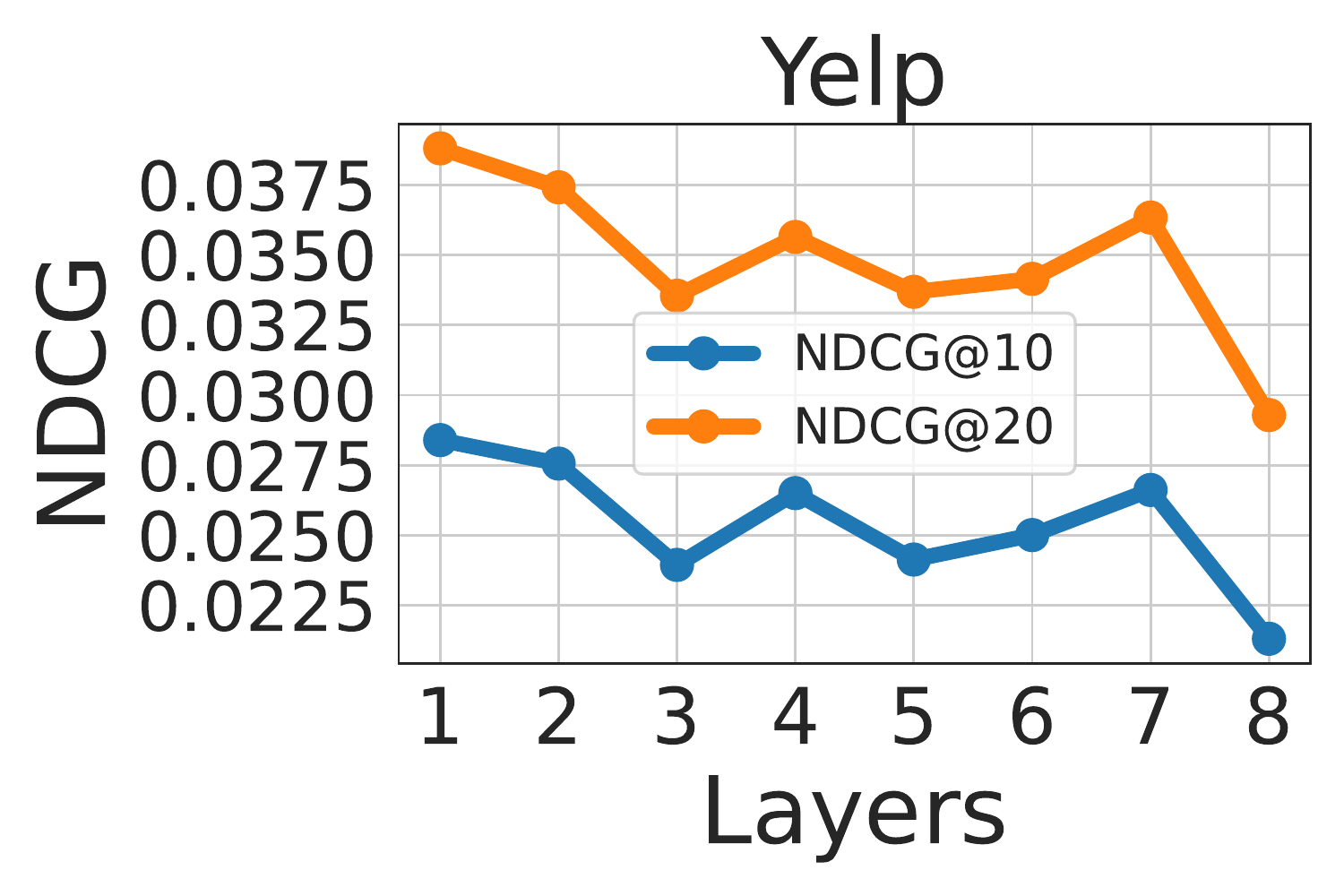}
     \end{subfigure}
        \caption{NDCG change with the increase of layers}
        \label{fig:ndcg layer}
\end{figure*}

In this experiment, we keep all the other hyper-parameters fixed, and observe the performance change with the increasing of convolution layer numbers. Experiment results on Recall@10 are shown in Fig.~\ref{fig:recall layer} and NDCG@10 are shown in Fig.~\ref{fig:ndcg layer}. We can observe that both the metrics on $4$ datasets show a downward trend. Although there are occasional increases, \modelname always achieves the best performance when the number of layers is $1$. \modelname transforms the knowledge graph into direct edges between items, which enables a direct message-passing between item pairs by using just $1$ single graph convolution. With more layers, the over-smoothing problem~\cite{zhou2020towards} exacerbates in \modelname because it builds lots of edges between items. One node will be smoothed by much more neighbors with the constructed edges. The results suggest we do not need to increase the complexity by stacking more convolution layers when utilizing \modelname. A simple $1$ layer convolution can achieve the best performance.

\subsubsection{Channel combination methods}

\begin{figure*}[htbp]
     \centering
     \begin{subfigure}[b]{0.22\textwidth}
         \centering
         \includegraphics[width=\textwidth]{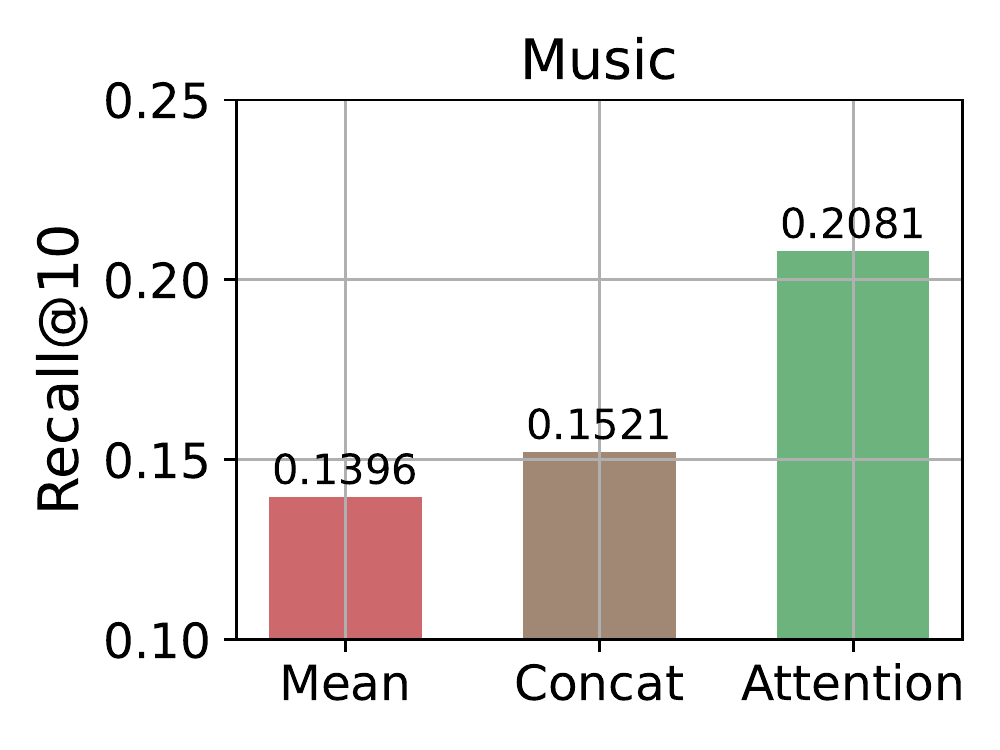}
     \end{subfigure}
     \hfill
     \begin{subfigure}[b]{0.22\textwidth}
         \centering
         \includegraphics[width=\textwidth]{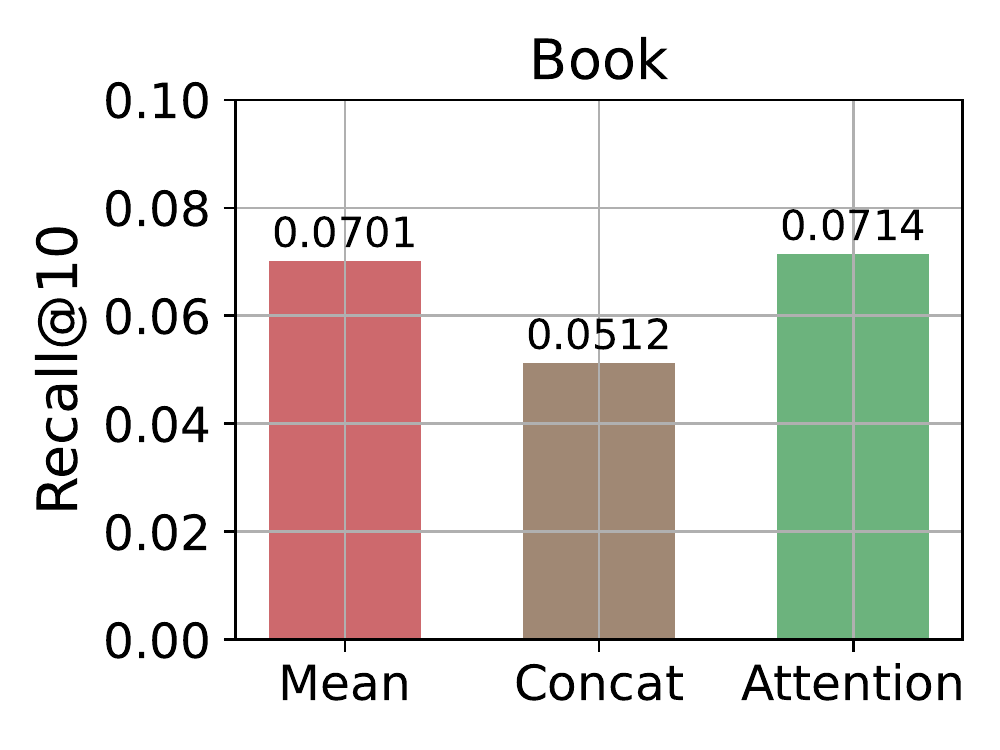}
     \end{subfigure}
     \hfill
     \begin{subfigure}[b]{0.22\textwidth}
         \centering
         \includegraphics[width=\textwidth]{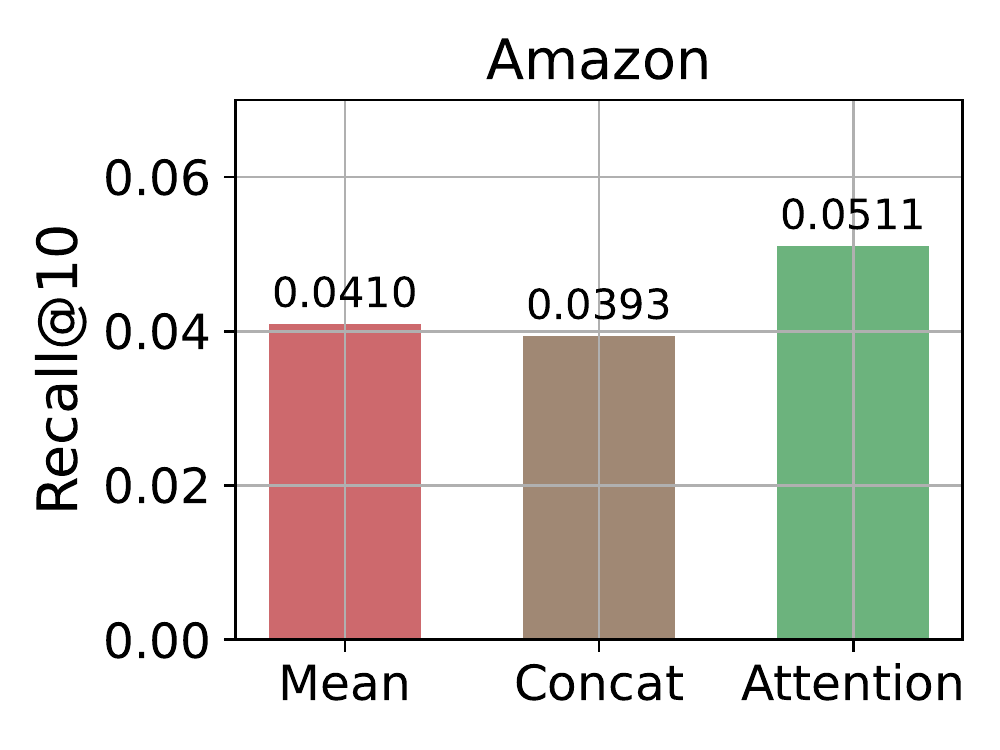}
     \end{subfigure}
     \hfill
     \begin{subfigure}[b]{0.22\textwidth}
         \centering
         \includegraphics[width=\textwidth]{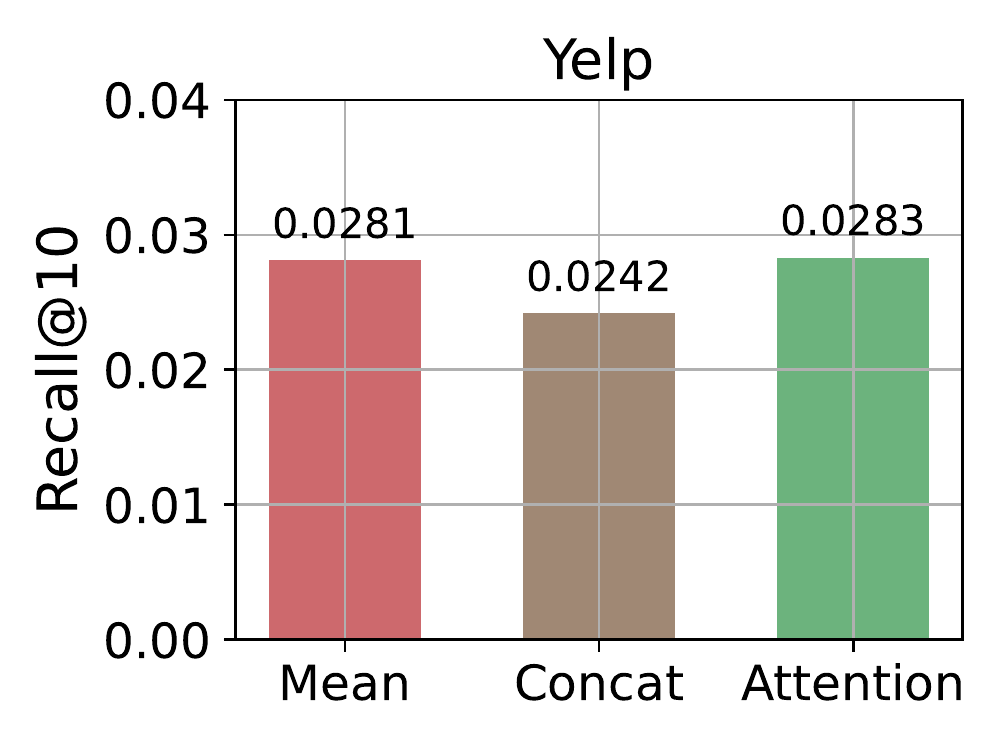}
     \end{subfigure}
        \caption{Recall change with different kinds of combination}
        \label{fig:recall combination}
\end{figure*}

\begin{figure*}[htbp]
     \centering
     \begin{subfigure}[b]{0.22\textwidth}
         \centering
         \includegraphics[width=\textwidth]{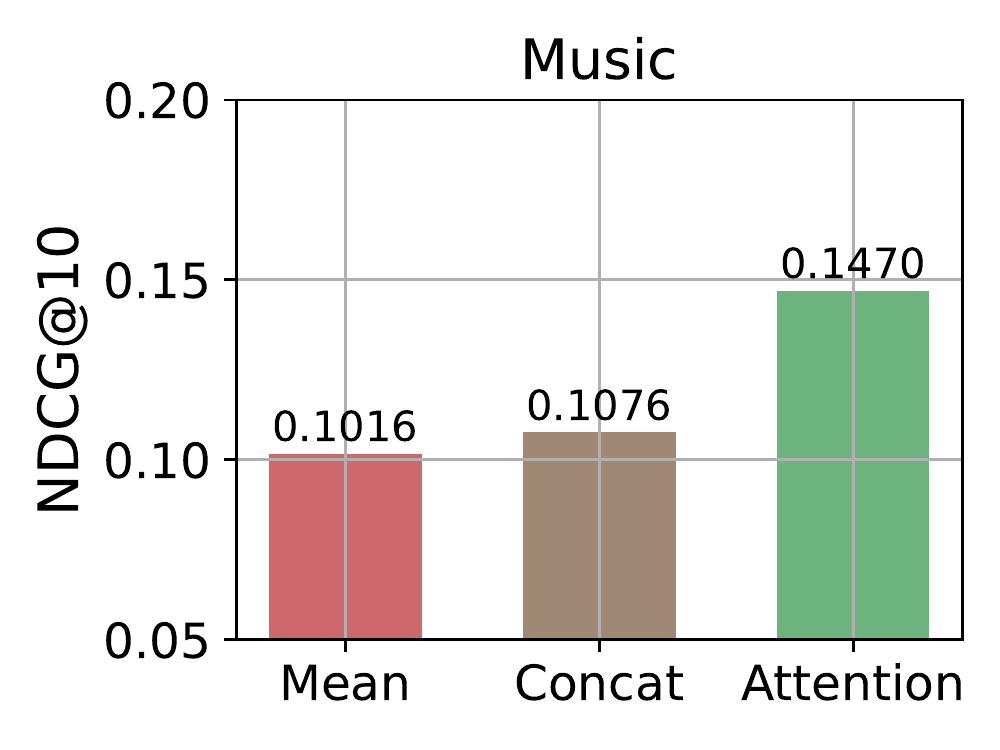}
     \end{subfigure}
     \hfill
     \begin{subfigure}[b]{0.22\textwidth}
         \centering
         \includegraphics[width=\textwidth]{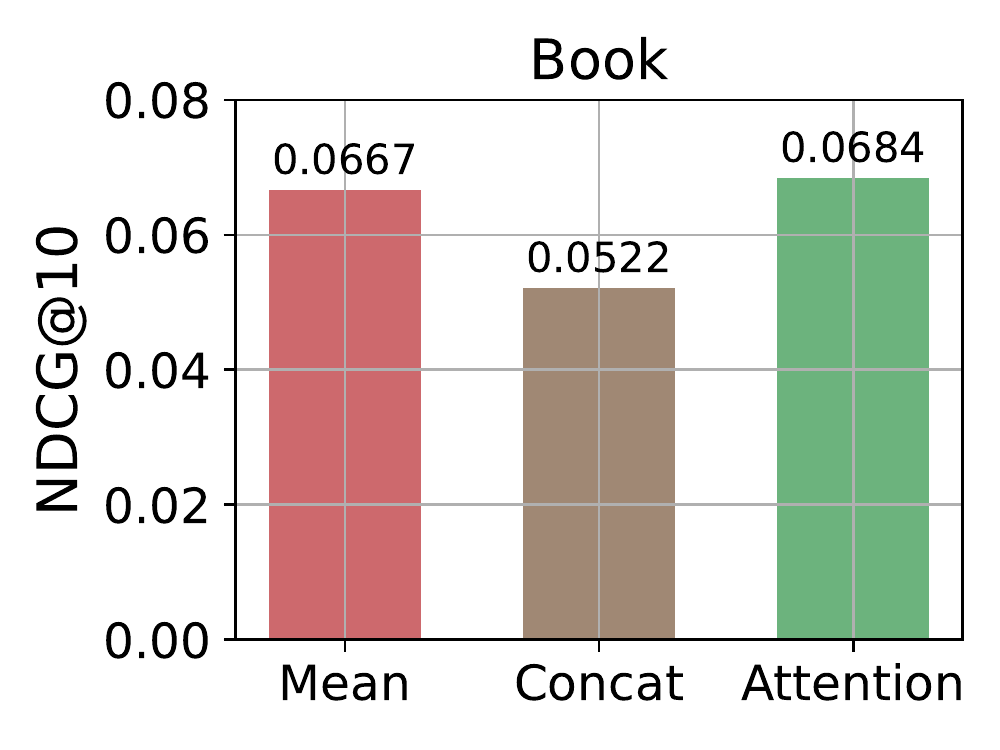}
     \end{subfigure}
     \hfill
     \begin{subfigure}[b]{0.22\textwidth}
         \centering
         \includegraphics[width=\textwidth]{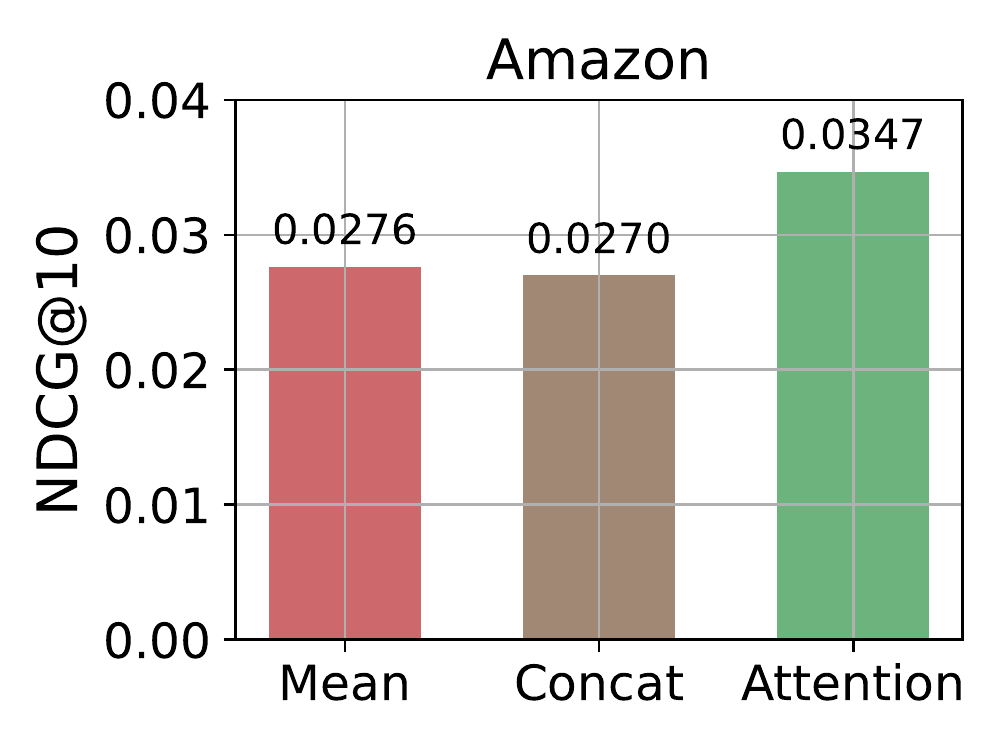}
     \end{subfigure}
     \hfill
     \begin{subfigure}[b]{0.22\textwidth}
         \centering
         \includegraphics[width=\textwidth]{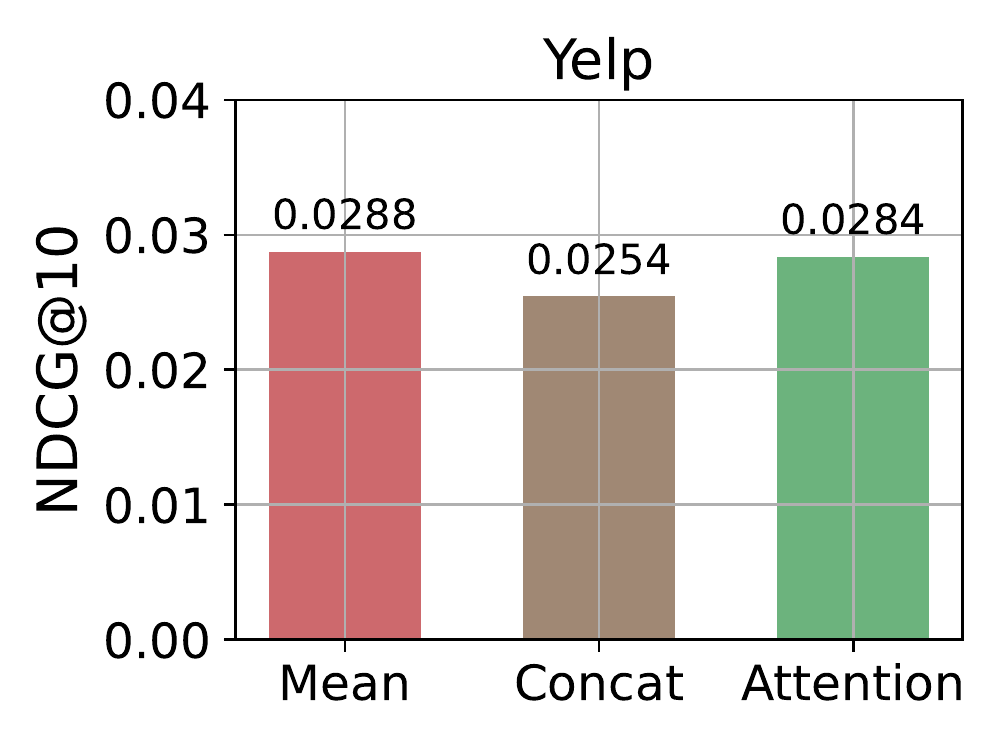}
     \end{subfigure}

        \caption{NDCG change with different kinds of combination}
        \label{fig:ndcg combination}
\end{figure*}

This experiment tests the channel combination methods. We keep all the other hyper-parameters fixed and only change the channel attention module to see the influence on performance. We show the changes on Recall@10 in Fig.~\ref{fig:recall combination} and NDCG@10 in Fig.~\ref{fig:ndcg combination}. In both figures, ``Mean" indicates we combine different embedding tables by a mean pooling layer, and ``Concat" indicates we first concat all the embedding tables and transform them to the previous embedding size by a linear layer. From the experiment results, we can observe that ``Attention" always achieves the best performance except for the NDCG@10 on Yelp dataset. On Music and Amazon dataset, ``Attention" surpasses the other two methods by a large margin. It validates the channel attention design in \modelname, and the attention can effectively fuse the embedding table learned from different Collaborative Meta-KGs.

\section{Related Work}
In this section, we introduce the related work of \modelname, which includes Meta Path/Graph learning and KG enhanced recommender system.

\subsection{Meta Path/Graph Learning}
Meta Path/Graph is a kind of powerful learning method on graph data, and it has attracted much research attention in recent years~\cite{chang2022megnn,li2021leveraging}.
To evaluate the relevance of different-typed objects, Shi et al.~\cite{shi2014hetesim} proposed HeteSim to measure the relevance of any object pair under an arbitrary meta path. Meta path-based methods are widely used on graph embedding. Meta-path2vec~\cite{dong2017metapath2vec} designed a meta-path-based random walk and utilized skip-gram to perform graph embedding. Sun et al.~\cite{sun2018joint} proposed meta-graph-based network embedding models, which simultaneously considered the hidden relations of all meta information of a meta-graph. Qiu et al.~\cite{qiu2018network} provided the theoretical connections between skip-gram-based network embedding algorithms and the theory of graph Laplacian and presented the NetMF method as well as its approximation algorithm for computing network embedding. Fan et al.~\cite{fan2018gotcha} proposed a HIN embedding model metagraph2vec to learn the low-dimensional representations for the nodes in HIN, where both the HIN structures and semantics are maximally preserved for malware detection. Fan et al.~\cite{fan2020metagraph} also introduced an attributed heterogeneous information network (AHIN) to model the rich semantics and complex relations among multi-typed entities and designed different metagraphs to formulate the relatedness between buyers and products. The HeCo proposed by Wang et al.~\cite{wang2021self} employed network schema and meta-path views to collaboratively supervise each other, moreover, a view mask mechanism was designed to further enhance the contrastive performance. 

Unlike previous works, we design meta graphs, particularly for knowledge graph enhanced recommendation. We propose Collaborative Meta Knowledge Graph that explicitly utilizes information from both knowledge graph and user-item interactions. The construction method is also not limited to paths or graph structures, which is flexible enough to encode different kinds of prior knowledge.

\subsection{KG Enhanced Recommender System}

KG contains rich entity and relationship information, which can be used as auxiliary information to supplement the relationship between users and items, thus making the recommender system more effective, accurate, and explainable~\cite{guo2020survey}. KG enhanced recommender system also receives much research attention~\cite{zhou2020improving,chen2020jointly,wang2019multi}. 
KTUP~\cite{cao2019unifying} and CFKG~\cite{ai2018learning} jointly learn recommendation and knowledge graph completion tasks by the shared embedding table.
KGCN~\cite{kgcn} is the pioneering work to take the advantage of GNN over knowledge graph to obtain an informative item embedding table. KGNN-LS~\cite{KGNNLS} learns the influence of the knowledge graph on users and transforms KG into a user-specific weighted graph before GNN aggregation.
Wang et al.~\cite{kgat} creatively propose the KGAT method, using two designs of recursive embedding propagation and attention-based aggregation to fully use the high-order information in KG to achieve the purpose of enhanced recommendation. 
Then they propose the KGIN~\cite{kgin} method, which provides a new aggregation scheme to extract useful information about user intent and encode them into user and item representations. 
ATBRG~\cite{feng2020atbrg} generates subgraphs for specific user-item pairs and uses a the relation-aware extractor layer to extract informative relations for aggregation. DSKReG~\cite{wang2021dskreg} samples on useful relations over knowledge graph by gumble softmax, and reconstructs a preference aware aggregation graph before GNN.

All previous researches rely on models to learn informative information from the complex knowledge graph with multiple kinds of relations. Unlike them, \modelname utilizes prior meta knowledge over knowledge graphs to construct different meta graphs to enhance recommendation. The reliable human-defined meta-knowledge decreases the noise and enables a simple model to achieve the best performance.

\section{Conclusion and Future Work}
In this paper, we research utilizing meta knowledge to enhance the KG-based recommender system. We propose constructing Collaborative Meta Knowledge Graphs to use prior meta-knowledge on the knowledge graphs and user-item interaction graphs. We make the construction based on different kinds of meta-knowledge and present \modelname that can effectively utilize knowledge graph to enhance recommendation. Experiment results on $4$ real-world datasets validate the effectiveness of \modelname. Cold-start experiment also shows \modelname can effectively utilize knowledge graph information to alleviate the cold-start problem. This work shows that the explicit usage of prior meta knowledge can benefit the KG-enhanced recommendation.

As for future work, we point out $2$ research directions. 1) Research on fusing prior meta-knowledge on other models. It is better to use meta knowledge to provide explicit signals when the relations are too complex for the model to learn.
2) Exploring how to construct explicit meta knowledge into deep learning models. In this paper, we propose the construction of Collaborative Meta-KG to encode meta-knowledge as edges between items. Researchers can also explore other methods to encode the meta knowledge in different scenarios.

\section{Acknowledgements}
Shen Wang, Liangwei Yang and Philip S. Yu are supported in part by NSF under grants III-1763325, III-1909323, III-2106758, SaTC-1930941. Jibing Gong, Shaojie Zheng, Shuying Du are funded in Hebei Natural Science Foundation of China under grant F2022203072.


\bibliographystyle{IEEEtran}
\bibliography{IEEEfull}

\end{document}